\documentclass[a4,useAMS,usenatbib,usegraphicx]{mn2e}

\def\aaps{A\&AS}
\def\apj{ApJ}
\def\apjl{ApJL}
\def\mnras{MNRAS}
\def\pasp{PASP}
\def\aj{AJ}
\def\araa{ARA\&A}
\def\aap{A\&A}
\def\apjs{ApJS}
\def\nat{Nature}
\def\pasj{PASJ}

   \title[The strong transformation of spiral galaxies infalling into massive clusters at $z\approx$0.2]{The strong transformation of spiral galaxies infalling into massive clusters at $z\approx$0.2}

  \author[L. Cortese et al.]{L. Cortese$^1$, D. Marcillac$^2$, J. Richard$^3$, H. Bravo-Alfaro$^4$, J.-P. Kneib$^5$, 
  G. Rieke$^2$, \newauthor G. Covone$^6$, E. Egami$^2$, J. Rigby$^2$, O. Czoske$^7$, J. Davies$^1$\\
  $^1$ School of Physics and Astronomy, Cardiff University, Cardiff CF24 3AA, UK\\
  $^2$ Steward Observatory, University of Arizona, 933, N. Cherry Avenue, Tucson, AZ 85721, USA\\
  $^3$ Caltech, Astronomy 105-24, 91125 Pasadena CA, USA\\
  $^4$ Departamento de Astronomia, Universidad de Guanajuato, Apdo. Postal 144, 36000 Guanajuato, Mexico\\
  $^5$ Laboratoire d'Astrophysique de Marseille, Traverse du Siphon, BP8 13376 Marseille, France\\
  $^6$ INAF-Osservatorio Astronomico di Capodimonte, Napoli, Italy\\
  $^7$ Argelander-Institut f\"ur Astronomie, Universit\"at Bonn, Auf dem H\"ugel 71, 53121 Bonn, Germany\\
    }
 
\begin{document}
\date{Accepted 0000 December 00.
      Received 0000 December 00;
      in original form }

\maketitle

\label{firstpage}

\begin{abstract}

 We describe two peculiar galaxies falling into 
 the massive galaxy clusters Abell 1689 (z$\approx$0.18) and 2667 (z$\approx$0.23) respectively. 
 Hubble Space Telescope images show extraordinary trails composed of bright 
 blue knots (-16.5$<M<$-11.5 mag) and stellar streams associated with each of these systems.  
Combining optical, near and mid-infrared and radio observations we prove that 
while both galaxies show similar extended trails of star-forming 
knots, their recent star formation histories are different. 
One ($\approx \rm L^{*}$) is experiencing a strong burst of star formation, appearing as a rare 
 example of a luminous infrared cluster galaxy.
 In comparison, the other ($\approx \rm 0.1 L^{*}$) has recently ceased 
 its star formation activity.
 Our model suggests that the morphologies and star 
 formation in these galaxies have been influenced by 
 the combined action of tidal interaction (likely with the cluster potential) 
and of ram pressure with the intracluster medium. 
 These results can be used to gain more insights to the origin of S0s, dwarf and ultra-compact dwarf (UCD) cluster galaxies.

\end{abstract}


\section{Introduction}
Clusters are natural laboratories to study the effects of environment on the 
evolution of galaxies \citep{dressler80}. 
A plethora of evidence shows that the properties of 
late-type galaxies depend strongly on environment: besides the well known morphology-density 
relation \citep{dressler80,whitmore}, in local clusters ($z\leq$0.03) 
spiral galaxies are deficient in neutral hydrogen 
\citep{giova85,cayatte90,hector1} and have lower star formation activity than 
galaxies of similar type and size in low density environments \citep{lewis02,gomez03,ha06}. 
Various physical mechanisms have been proposed to explain the different evolution of 
late type spirals in clusters and in 
the field. In general, they invoke either dynamical interactions of cluster galaxies with the 
hot intracluster medium (ICM, \citealp{GUNG72,LARS80}), or gravitational interactions 
with nearby companions \citep{merritt83}, with the potential of the cluster \citep{byrd1990,valluri93}, 
or with a combination of these two \citep{harrassment}.
Interactions with the ICM are likely to be the dominant process at the
present epoch and can account for the truncation of the gas disks in members of several 
local clusters (see \citealp{review} and references therein). 
However, ram-pressure cannot produce the strong morphology-density relation \citep{dressler80,whitmore}, 
nor can it thicken the disk of a spiral galaxy and transform it into an S0 
(i.e. \citealp{hinz03,christlein04,n4569}).\\
This apparent contradiction could be solved if the structures form hierarchically;
that is, if galaxy clusters form not by accreting individual galaxies randomly from the field, 
but rather by the infall of less massive groups along filaments. 
These infalling groups have velocity dispersions that are much smaller than that of the cluster 
as a whole, permitting the slow, strong interactions normally 
associated with field galaxies \citep{FUJI04,mihos04,dress04car,big}.
Therefore a plausible evolutionary history would take into account that environmental conditions 
and the physical properties of galaxies are changed significantly during cosmic time, changing the 
influence of various physical mechanisms on the evolution of galaxies \citep{review}. 
However, this hypothesis is far from confirmed, since we lack detailed 
understanding of the range of environmental effects that act as a function of 
the age of the Universe \citep{dress04car}.
Although star-forming galaxies in clusters at intermediate redshift appear more 
disturbed \citep{oemler97,couch98} 
and have higher star formation activity \citep{butcher1,butcher2,fadda00} than local disk 
galaxies, it is still an open question which mechanisms are at play and how they influence 
the evolutionary history of cluster galaxies \citep{balogh99,dressler99,treu03}.
To solve this riddle, we need to observe galaxies that physical circumstances 
and chance have revealed in rare moments of transformation. 
These peculiar systems can be used to probe different environmental effects 
and to constrain models of the evolutionary history of galaxies. 
Much of our knowledge on the evolution of nearby galaxies in both groups \citep{duc00,IGLV01,sulentic01} and clusters 
\citep{n4438,big,GAVB01,kenney95,vollmer01,vollmer04} has in fact come from studying of 
such systems. Unfortunately this information 
is difficult to extend to high redshift 
because both clusters and galaxies have evolved significantly: clusters were less 
relaxed \citep{jeltema05} and galaxies had higher gas content. Therefore,
the effects of the same environmental mechanisms could depend strongly on 
the age of the Universe.\\
In this paper, we present a multiwavelength analysis of two peculiar galaxies 
(hereafter referred as 235144-260358 and 131124-012040) falling into the centers 
of the massive clusters Abell 2667 ($z\approx$0.23) and Abell 1689 ($z\approx$0.18).
Both these systems are associated with extended 
trails of bright blue knots and diffuse wisps and filaments of young stars, 
features observed so far only in one other galaxy at similar redshift \citep{owen06}. 
These two objects have been serendipitously detected by looking at the
WFPC2 and ACS images of massive clusters at z$\approx$0.2.  The sample of clusters
consist of the 10 clusters discussed in \cite{smith_clust05} plus A2667, A2390 and
A1689. We therefore found 2 galaxies with extended trails within 13 studied 
clusters all located at 0.175$<z<$0.25, suggesting that we are observing a very short 
snapshot of a critical stage in the evolution of these cluster galaxies.\\ 
Because these two systems have 
significantly different optical luminosities ($\approx\rm  L^{*}$ and $\approx$0.1$\rm L^{*}$) 
but are at similar distances from the cores of clusters of similar mass, 
they represent an interesting case for a comparison of the effects of similar 
environments on different-sized galaxies.\\
Throughout this paper we assume a cosmology
with ${\Omega}_m$ = 0.3, ${\Omega}_{\lambda}$ = 0.7 and $H_0$ =
70 km/s Mpc$^{-1}$, implying a distance modulus of 39.74 (40.33) mag and a linear 
scale of 3.16 (3.71) kpc/arcsec for A1689 (A2667).

\section{The Data}
\subsection{Optical Photometry}
\begin{figure*}
\includegraphics[width=17cm] {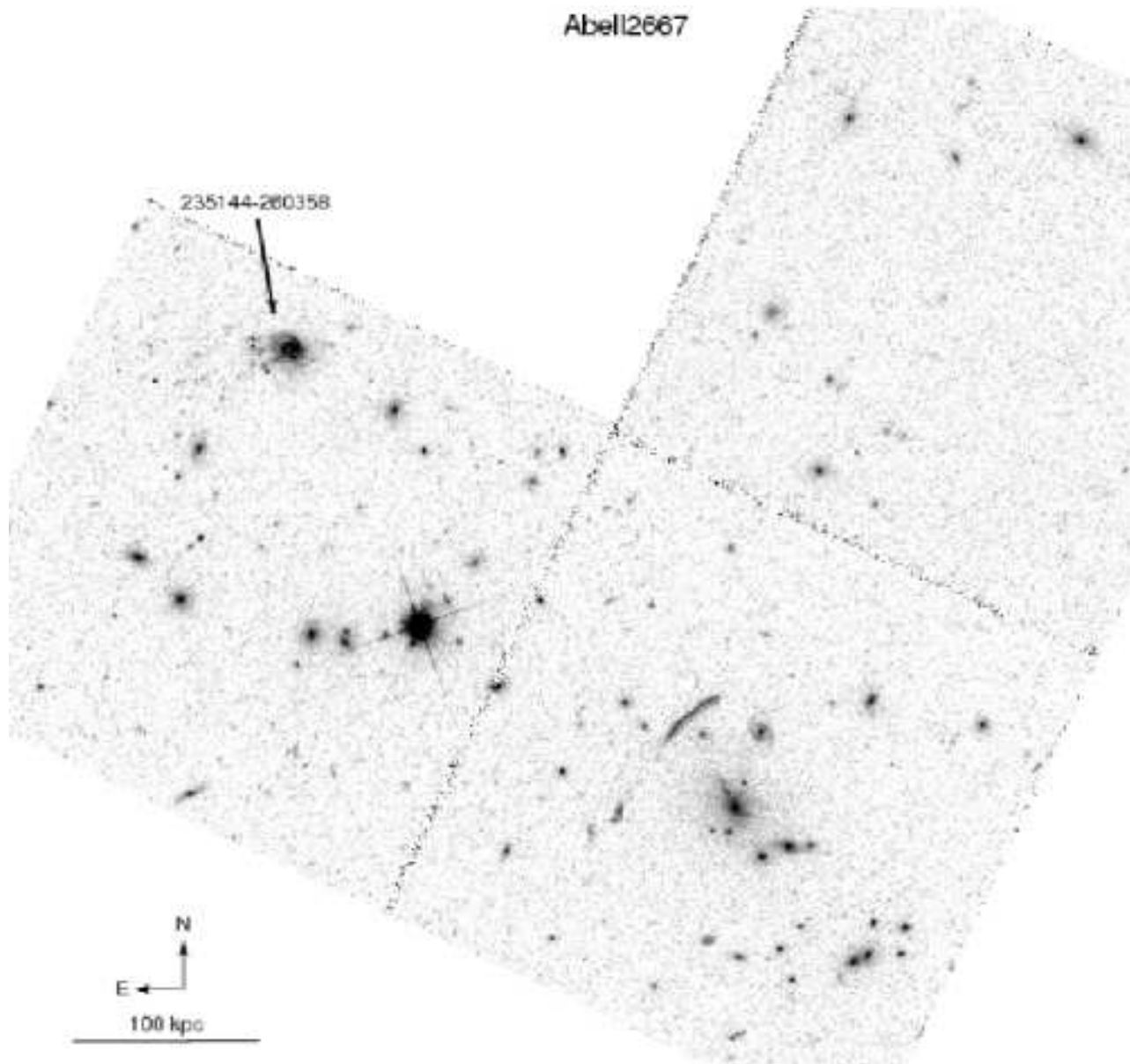}
\caption{HST-WFPC2 F450W image of Abell 2667. The position of 235144-260358 is indicated.}
\label{whole2667}
\end{figure*}

\begin{figure*}
\includegraphics[width=17cm] {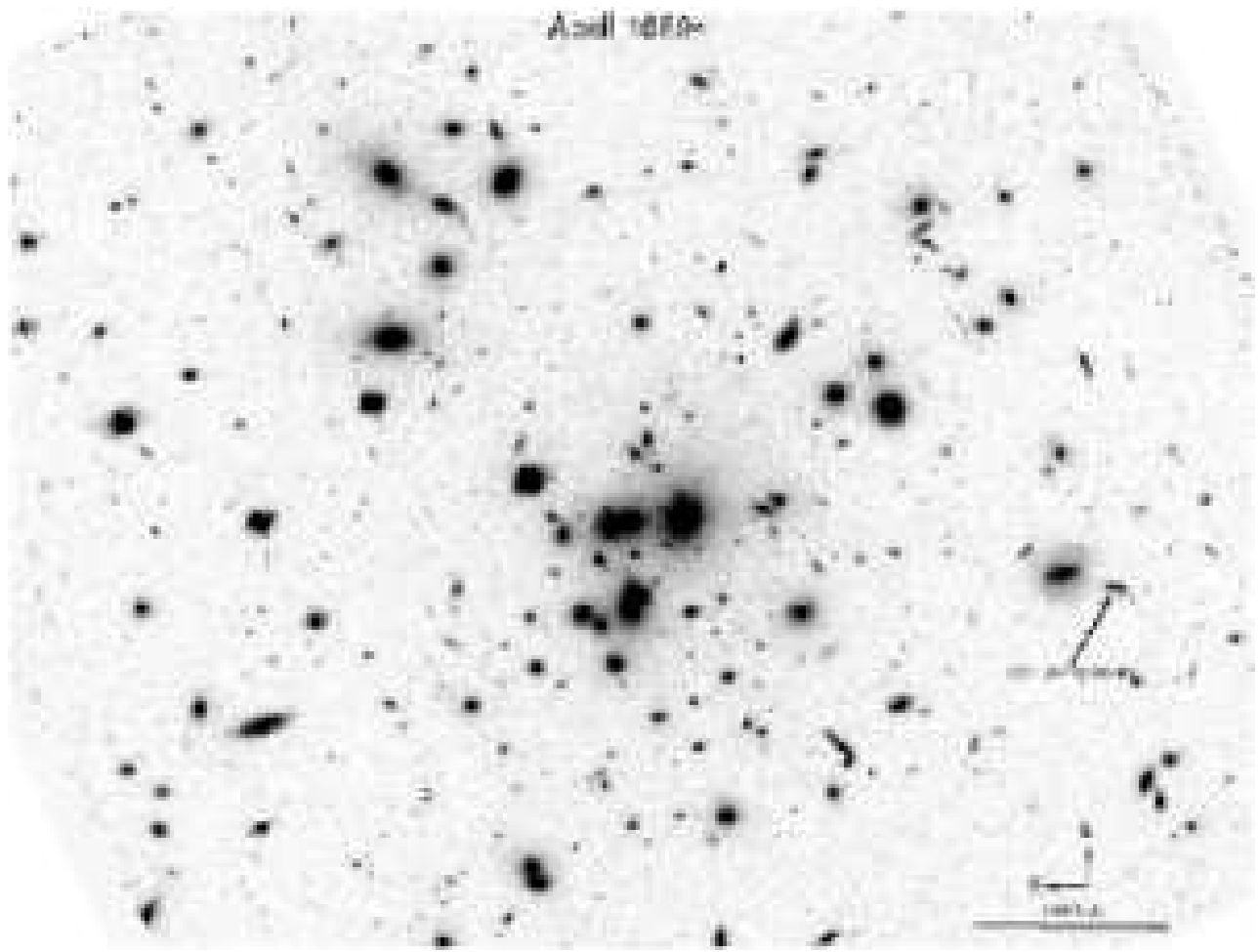}
\caption{HST-ACS F475W image of Abell 1689. The position of 131124-012040 is indicated.}
\label{whole1689}
\end{figure*}

The optical photometric data for this paper are extracted from 
deep $HST$ images of Abell 2667 and Abell 1689.
Abell 2667 was observed on 2001 October, using the Wide Field Planetary 
Camera (WFPC2) for total exposures of 1200 seconds through the F450W filter,
and 4000 seconds in F606W and F814W (see Fig. \ref{whole2667} and Fig.\ref{colimage}) \citep{covone05}.
The 3 $\sigma$ detection limit for point sources is $\approx$ 26.00, 26.00 and 25.00 mag in 
the F450W, F606W, and F814W bands, respectively.
Deep observations of Abell 1689 were obtained from the ACS Guaranteed Time 
observations in 2002 June (see Fig. \ref{whole1689} and Fig.\ref{colimage}).
A total of 20 orbits ($\approx$13.2 h) were taken in the three passbands F475W, F625W and 
F850LP \citep{broad05}.
The 3 $\sigma$ detection limit for point sources is $\approx$ 29.70, 29.20 and 28.30 mag in the filters 
F475W, F625W, F850LP respectively.\\   
We used SExtractor \citep{sex} to detect and analyze the sources.
For source detection, we used an image averaging the three band-passes; magnitudes 
were then determined from aperture photometry on the separate 
images for each filter. All magnitudes are in the VEGAMAG systems. 
No correction for Galactic extinction was performed ($A_{V}\leq$0.07 mag). 
Surface brightness profiles for these galaxies were computed using 
the task {\it ellipse} within IRAF. 
The ellipticity and position angle were determined using the I band images 
following the procedure of  \cite{gav00}.
The disturbed morphologies of 235144-260358 in A2667  could possibly affect the shape of the surface brightness 
profiles at large radii (in particular at shorter wavelengths), but not in the central regions where both the objects 
still present a reasonably symmetrical shape.  This is not the case of the edge-on spiral 131124-012040 which does 
not show strong asymmetries within the optical radius.

\subsection{Near Infrared Photometry}
Near-infrared H band observations for Abell 2667 and Abell 1689 were 
obtained with ISAAC on the Very Large Telescope (VLT) in the 
spring of 2003 (ESO Programs 071.A-0428, PI. Kneib
and 067.A-0095 PI. Knudsen), under photometric sky conditions with a mean seeing of 
$\approx$0.41 arcsec and $\approx$0.58 arcsec respectively. 
The total exposure time of 6529s for each cluster corresponds 
to a 5$\sigma$ detection limit
for point sources of $\approx$24.6 mag.
All these observations have been reduced as described 
in \cite{richard06}.

\subsection{Mid Infrared Photometry}
Spitzer imaging observations of Abell 2667 and Abell 1689 were obtained as part 
of the GTO Lensing Cluster Survey (program 83, PI G.~Rieke).
MIPS \citep{mips} 24$\rm \mu m$ images were obtained in photometry mode, with a total
exposure time of $\approx$2700s. The data were processed and
mosaicked following the procedures described in \cite{egami06}. 
Point source extraction and photometry were performed using DAOPHOT \citep{daophot} as described 
in \cite{papovich04}. A PSF was constructed from the best-measured 30 point sources in the field; the Tiny Tim model of the 24$\rm \mu$m 
PSF \citep{krist} was used to compute the aperture 
correction. 131124-012040 in Abell 1689 is not detected in MIPS images. We derived a 3 $\sigma$ limit using 
a photometry aperture radius of 6 arcsec, a sky annulus between 6 and 13 arcsec and an aperture correction of 1.698\\
IRAC \citep{irac} four-bands (3.6, 4.5, 5.8 and 8.0$\rm \mu m$) imaging was also obtained, with a total exposure time of
2400s per band for each cluster.  Basic calibrated data were combined using 
a custom IDL mosaicking routine.
For A2667 photometry was performed within apertures of radius $\approx$ 8.3 arcsec
and no aperture corrections were applied because they are small with this large an extraction aperture.
On the contrary in A1689 photometry was performed within a smaller aperture to 
avoid light contamination from nearby sources. We adopted a radius $\approx$ 2.4 arcsec, sky annulus between 
2.4 and 7.2 arcsec and aperture corrections of 1.213, 1.234, 1.379 and 1.584 at 3.6, 4.5, 5.8 and 8.0$\rm \mu m$ 
respectively.  

\subsection{Radio continuum observations}
We obtained a 20\,cm radio continuum measurement of ${\approx}$1.4\,mJy for the 
galaxy in A2667 from the 1.4\,GHz NVSS continuum survey. As this survey offers  rather poor spatial resolution ($\approx$45
arcsec), we also constructed a 20\,cm map using higher
resolution data from the NRAO VLA data-archive.
Two different observations are available on A\,2667 at this frequency:
1) it was observed in October 2001 for 3590 seconds in
correlator mode 4, with a bandwidth of 25 MHz, and using 
the CD-configuration (due to the low declination); and 2.) 
a 3620 second observation was obtained in September 2002
with the same correlator mode and bandwidth, but in the BC
configuration. We applied
standard VLA calibration and imaging procedures with AIPS. 
We then combined the two data sets in the UV-plane using DBCON. 
Images were generated with the task IMAGR and a weighting 
option {\it robust}=0, producing a map 
intermediate between natural and uniform weighting.  
After CLEANing, the resulting continuum map has 
a beam size of 16.7 $\times$ 13.1 arcsec and an average rms noise of
0.12 mJy/beam.
This map is shown in Fig.\ref{radio}, superposed on the HST 
image. \\
The cluster A\,1689 was observed with the VLA for a total of 17405
seconds in the A-configuration in November 2000 and March 2002. 
Combining the data sets using natural weighting (to improve 
the sensitivity) we produced a 
map with a beam size of 2.1 $\times$ 1.6 arcsec, and an 
average rms of 0.15mJy/beam.
No emission is detected at the position of the infalling galaxy. 
Taking a conservative detection threshold 
of 6$\sigma$ we estimate an upper limit of 0.90\,mJy for the 20\,cm 
radio continuum flux.

\subsection{Optical spectroscopy}
We obtained optical spectroscopy for 131124-012040 in Abell 1689 as part of a wide 
field ($\approx 30\arcmin\times30\arcmin$) spectroscopic survey of the whole cluster (\citealp{Czoske2004}, Czoske 
et al.\  2007, in preparation) using VIMOS on the VLT (Program 71.A-3065, P.I. Czoske). 
The LR-Blue grism was used, which provided a resolution
of $R\approx 200$ over a wavelength range from 3750\,\AA\ to 6750\,\AA.
The dispersion was 5.35\,\AA\ per pixel. 
We obtained three exposures of 840~seconds, for a total of 
42 minutes.
The data were reduced using VIPGI \citep{Scodeggio2005} on site at
IASF in Milano. The reduction involved bias subtraction, identification of
the spectrum on the two-dimensional spectral image, interactive wavelength
calibration from observations of arc spectra and optimal extraction using
the method of \citet{Horne1986}. The spectrum has been flux-calibrated
from observations of a spectrophotometric standard star, Feige~110.\\ 
In addition we observed A2667 and A1689 in June 2006 with the LRIS instrument \citep{Oke95} on Keck I.
A 5600 \AA\ dichroic separated the red channel of the instrument, equipped with a 400 l\ mm$^{-1}$ 
grating blazed at 8500 \AA, from the blue channel, equipped with a 600 l\ mm$^{-1}$ grism blazed at 4000 \AA . 
This setting covers the wavelength range $3300-9200$ \AA\ with a dispersion of 0.6/1.8 \AA\ and 
a resolution of 4.3/6.3 \AA\ in the blue/red channel, respectively.
On June 29, a 175\arcsec-long and 1\arcsec-wide slit was aligned on the center 
of 235144-260358 in A2667 including two of its knots (K1 and K2, see Fig.\ref{specknots}), using a position angle of 92.7 East of North.
Two exposures of 900 s were obtained under $\approx 1.5$\arcsec seeing. 
On June 30, a 30-slits mask was used on A1689 in order to target multiple-imaged candidates at the cluster center. 
One blue knot associated with the disrupted galaxy was included in a 9\arcsec-long and 1\arcsec-wide slit from this mask (see Fig.\ref{specknots}). 
Four exposures of 1800 sec each have been obtained with an average seeing of 1.0\arcsec.
All these spectroscopic data were reduced using standard IRAF procedures for flat-fielding, wavelength and 
flux calibrations, sky subtraction and extraction of the spectra.\\
Under visual inspection of the spectra we carried out the measurement of the emission and absorption 
lines using the task SPLOT in IRAF. 
For the spectrum of 235144-260358 in Abell 2667 we de-blended the underlying absorption from the 
H$\beta$ emission lines as discussed in \cite{gavspectra}. We evaluated the Balmer decrement from the 
ratio H$\beta$/H$\alpha $ (assuming $T$=10 000 K and $n$=100 $\rm e/cm^3$, \citealp{osterb89}) and derived the corrected 
line fluxes, relative to H$\beta$, using  the de-reddening law of \cite{lequex}.
The observed H$\alpha$/H$\beta$ ratio for 235144-260358 is $\approx$3.65 implying a gas attenuation $A(H\alpha)\approx$0.56 mag and 
a stellar continuum attenuation $A(V)\approx$0.31 mag (assuming the Galactic extinction curve, \citealp{pei92}).

\subsection{X-ray Imaging}
We downloaded the 9.2 ks Chandra 
ACIS observation of A2667 from the public
archive and reduced it following the standard "threads" from CIAO data analysis software (Version 3.3)\footnote{http://cxc.harvard.edu/ciao/index.html}.  We searched the
exposure--corrected images for sources, using the task wavdetect with
angular wavlet scales from \cite{brandt01} and a
significance threshold of $1\times 10^{-7}$. No source is detected at the
position of 235144-260358. We measure an upper limit to the 2--8 keV flux of
$1.2\times10^{-14}~\rm erg~s^{-1}~cm^{-2}$.

\section{Results}

\subsection{Abell 2667}
\begin{figure*}
\includegraphics[width=17cm] {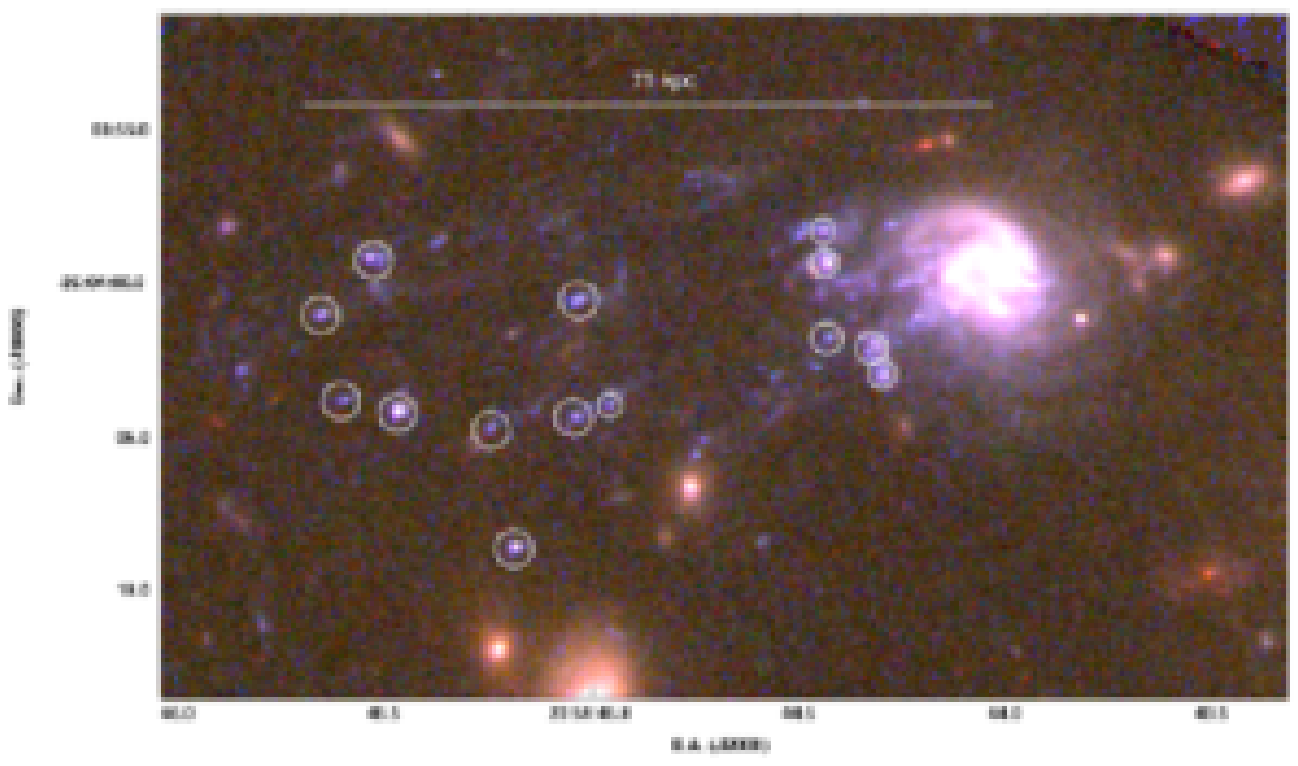}
\includegraphics[width=170mm] {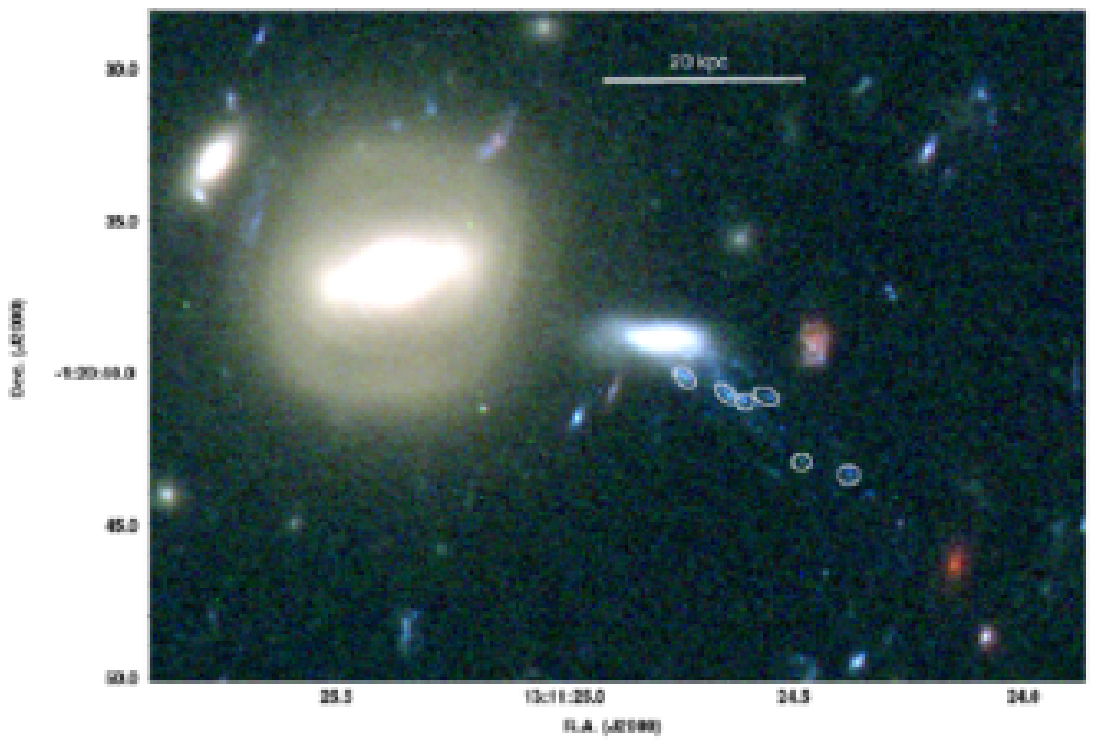}
\caption{RGB color image of 235144-260358 in Abell 2667 (upper panel) and of 131124-012040 in Abell 1689 (lower 
panel). The white circles indicate the knots studied in Sec. 3}
\label{colimage}
\end{figure*}

\begin{figure*}
\centering
\includegraphics[width=7.5cm]{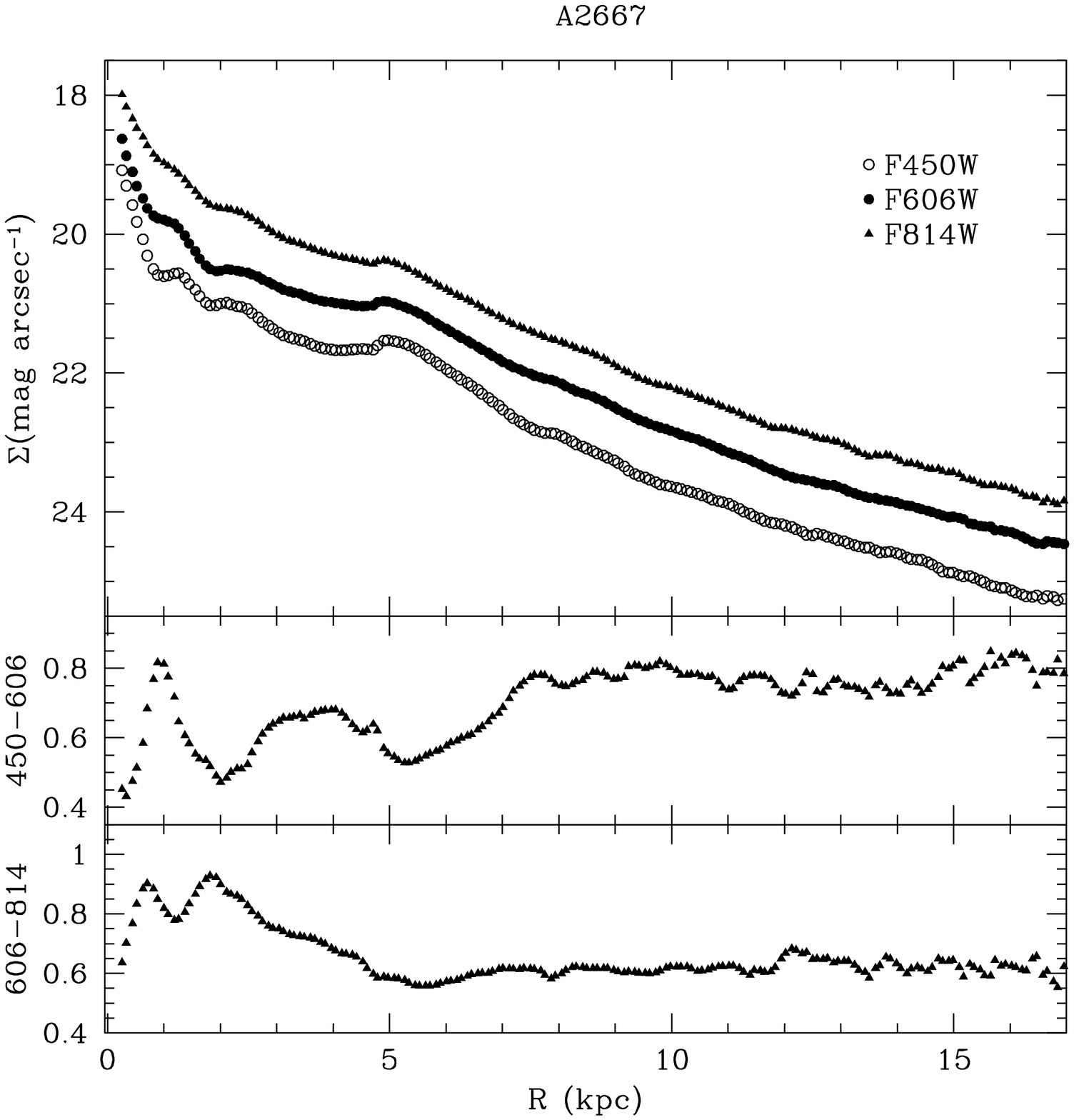}
\includegraphics[width=7.5cm]{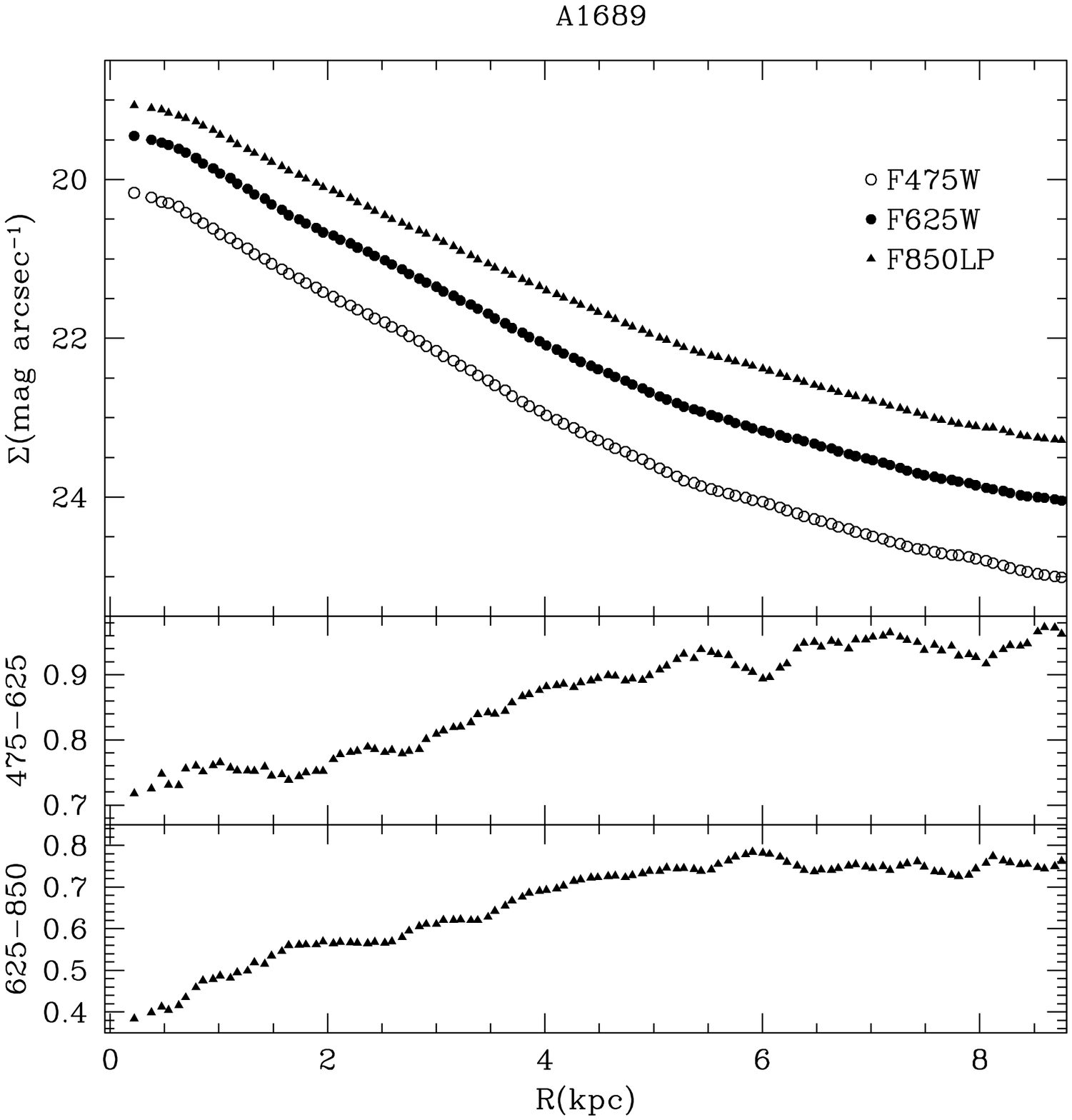}
\caption{The surface brightness and color profiles in the three HST bands for 235144-260358 in Abell 2667 (left) 131124-012040 in Abell 1689 (right).}
\label{colprofiles}
\end{figure*}

\begin{figure*}
\centering
\includegraphics[width=12.5cm]{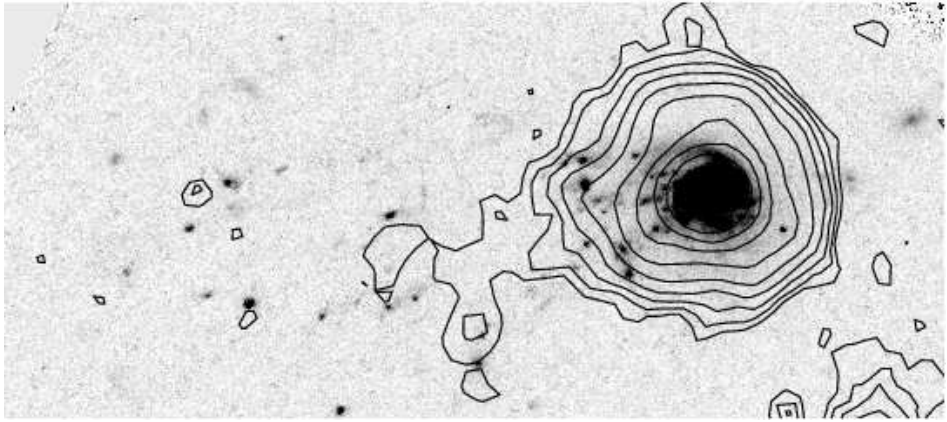}
\caption{HST F450W image of 235144-260358 in Abell 2667. The Spitzer 8 micron contours are superposed in black. 
The lower contour corresponds to a 3$\sigma$ noise level ($\approx0.23~\rm \mu J~arcsec^{-2}$).}
\label{8micron}
\end{figure*}

\begin{figure*}
\centering
\includegraphics[width=12.5cm]{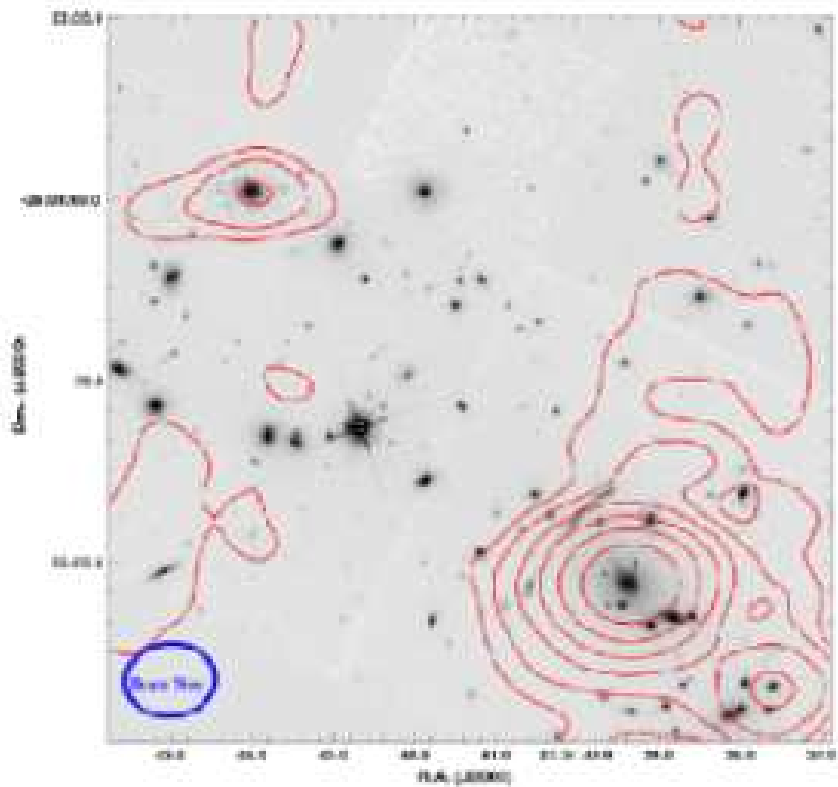}
\caption{The VLA 20cm radio contours (2,4,8, 16, 34 and 64 $\sigma$) of Abell 2667 superposed on the HST F450W image. }
\label{radio}
\end{figure*}

Fig. \ref{colimage} (upper panel) shows an RGB image of 235144-260358 in Abell 2667, 
and its global properties are summarized in Table \ref{tabgal}.
Its optical redshift is $z\approx0.2265$, lying in the low velocity tail of the 
velocity distribution of Abell 2667 (i.e. $\approx$830 $\rm km~s^{-1}$ lower than the mean cluster velocity; \citealp{covone05}). 
This face-on galaxy lies at a projected distance of $\approx$ 0.34 $h_{70}^{-1}$ Mpc from 
the cluster center (assumed to coincide with the position of the central CD galaxy of A2667, see 
Fig. \ref{whole2667}). This system is one of the brightest galaxies 
in the cluster  \citep{covone05}, with both optical and 
near infrared ($M_{F450W}\approx$-21.50, $M_{H}\approx$-24.50) absolute
magnitudes close to L$^{*}$ and a gas metallicity\footnote{The gas metallicity has been computed from the average 
of five different empirical determinations based on:
$R_{23}$ \citep{zaritsky94,mcg91}, $\rm [NII]\lambda6583/[OII]\lambda3727$
\citep{kewley02}, $\rm [NII]\lambda6583/H\alpha$ \citep{vanzee98} and
$\rm [OIII]\lambda5007/ [NII]\lambda6583$ \citep{dutil99}} of $12+log(O/H)\approx$9.0$\pm$0.1 (i.e. $\approx$1.4 solar metallicity).
From the HST images, 235144-260358 appears to be a late-type galaxy (see Fig.\ref{colimage}), as 
confirmed by its structural parameters (see Table \ref{tabgal}).
However this object is definitely not normal: it shows a disturbed morphology, 
with clear indications of stripping within its optical disk and a prominent one-armed 
spiral component as is typically observed in gravitationally perturbed 
systems \citep{vollmer03}. 
Moreover, there is a significant nuclear enhancement in the optical surface brightness profiles ($\approx$ 2 mag within 
the central kpc: see Fig.\ref{colprofiles}), suggesting that 
it is experiencing a nuclear burst of star formation. 
This spike is particularly evident in the F450W band, where the central regions cannot be fitted with a simple deVaucouleurs profile. \\
Spitzer observations of A2667 expand on the unusual properties of 235144-260358, since 
it is detected by both IRAC (3.6-8$\rm \mu m$, see also Fig.\ref{8micron}) and MIPS (24 $\rm \mu m$). 
At $z\approx$0.23, the 8 $\rm \mu$m emission is dominated by  a combination of 
the PAH bump ($\approx6.2\rm \mu$m) and very small grains 
continuum \citep{DBP90}, while the old stellar population dominates at shorter wavelengths.
The observed 8$\rm \mu m$/5.6 $\mu m$ flux ratio $\approx$ 6.3 
(corresponding to the rest frame flux ratio 6.3$\rm \mu m$/4.5 $\rm \mu m$) is consistent with 
the value observed in star forming galaxies \citep{dale05}, suggesting that the infrared emission 
is due to recent star formation activity.\\
We used the X-ray data for a second test of whether the infrared emission 
is due to a burst of star formation or to an active nucleus (AGN). 
Comparing the X-ray upper limit to the 24 $\rm \mu m$ flux density (see Table \ref{tabgal}) 
with the help of figure~1 of \cite{alonso04} confirms 
that this source is not AGN--dominated:  its
2--10 keV/24$\rm \mu m$ flux ratio is at least 4 times too 
low to lie within the range of typical AGN \footnote{Neither
of these tests can exclude the presence of a 
Compton thick AGN \citep{shi05}, but it is likely that the mid-infrared
output of such objects is dominated by star formation.}. Finally a significant 
contribution from an AGN is ruled out by the emission line ratios 
obtained from the optical spectrum: $\log([OIII]/H\beta)\approx$-0.45, $\log([NII]/H\alpha)\approx-0.33$ 
consistent with the values typically observed in star forming galaxies \citep{kewley01}.
We therefore used the 8 and 24$\rm \mu$m data to derive the total infrared luminosity, $L(IR)$, 
using the IR spectral energy distribution (SED)  
from the \cite{dale02} and \cite{chary01} library following the procedure described in \cite{marcillac06}. 
This method relies on the correlations between L(IR) vs. the luminosity at 7 $\mu m$ and $L(IR)$ vs. $L(24 \mu m)$ 
shown in \cite{chary01}. The SED templates were only used to interpolate at 8 and 24 $\mu m$.
The resulting total infrared luminosity of 235144-260358 $L(IR)\approx$3($\pm$0.25) $10^{11}~\rm L_{\odot}$, 
implies a current star formation rate $SFR\approx$ 53($\pm$4.3) $\rm M_{\odot}~yr^{-1}$ (using the relation of \citealp{kenn98}), 
consistent with a $SFR\approx$ 57 $\rm M_{\odot}~yr^{-1}$ obtained from VLA continuum observations and the relation of 
\cite{condon92}. This galaxy is a rare example of a luminous infrared galaxy (LIRG) in a dense cluster.\\
All the properties of 235144-260358 point out 
the peculiarity of its recent evolutionary history. 
However, the extended trails of bright blue knots, tracing its trajectory as
it falls into the cluster core, make it truly extraordinary.
A dozen such knots extend from the galaxy 
optical disk to a projected distance of $\approx$ 80 $h_{70}^{-1}$ kpc. 
Extended blue low surface 
brightness wisps and filaments lie along the same 
trail, supporting the hypothesis that all of these structures result 
from stripping (see Fig.\ref{colimage}). 
The knots have absolute F450W magnitudes in the range -16.80$<M_{F450W}<$-14.80 mag, 
typical of dwarf galaxies \citep{sandage85} and super star clusters \citep{larsen99}, and are barely resolved in the HST image 
implying an effective radius $r_{e}\leq$0.45 $h_{70}^{-1}$ kpc.\\
The radio contours shown in Fig.\ref{radio} appear elongated in the 
direction of the trail. A similar morphology seems also 
to be present in the Spitzer 8$\rm \mu m$ map shown in Fig.\ref{8micron}, 
which has the appearance of a head on the galaxy, with a tail tracing
the current star formation associated with the blue knots.
Moreover [OII] emission, not associated with any of the blue knots, extends from the galaxy for 
a total length of at least $\approx$50 kpc (see Fig. \ref{specknots}), suggesting the presence of diffuse ionized gas along the trails as already 
observed in nearby ram pressure stripped galaxies \citep{GAVB01,yoshida04,big}. \\
To constrain the ages of the blue knots, we compute the time evolution of 
the $F450W-F606W$ and $F606W-F814LP$ colors, using Starburst99 \citep{starburst}. 
We assume a Salpeter IMF, 
solar metallicity\footnote{We also tested a Kroupa IMF and stellar metallicities in the range 0.004$<Z<$0.02, 
but the evolutionary paths do not significantly vary from the ones shown in Fig.\ref{ccdiagram}}  and two different star formation histories: an instantaneous burst and continuous star formation.
For each star formation history we also compute a model including the contribution 
of strong emission lines. 
We redshifted the synthetic Starburst99 spectra to the cluster distance 
and used the synthetic photometry package SYNPHOT in IRAF to compute the 
model colors in the WFPC2 passbands. 
In Fig. \ref{ccdiagram}, we plot a color-color diagram for the bright knots 
(black circles) to compare with the theoretical evolutionary tracks (solid and 
dashed lines). We present only the colors for the brightest knots: i.e. detected in all three HST bands 
(see Fig.\ref{colimage}). 
The arrow shows the effect of attenuation by dust on the observed colors, assuming a Galactic attenuation 
curve \citep{schlegel98}.  
The models with emission lines (dashed lines) appear to fit the observed colors better than those 
without (solid lines). 
Most of the blue knots lie slightly below the modeled tracks for 
both an instantaneous burst and continuous star formation but are reasonably consistent 
with an age of the episode in the range $5<t<15$ Myr in the first case and $10<t<1000$ Myr in the second.
These values are probably upper limits since it is very likely that the star forming knots contain dust, 
as observed in extragalactic HII regions \citep{gerhardHII,corteseHII} and star 
forming dwarf galaxies \citep{bosellised,COdust05}.\\
Our spectra detect $\rm [OII]$ in emission in both the knots (K1 and K2) included on the slit (see Fig. \ref{specknots}) confirming  
that these systems are still forming stars. 
For K1 we also detected 
[OIII] and H$\alpha$ in emission ($z\approx0.227$). 
The H$\alpha$ flux is $f\approx1.6\times10^{-17}\rm~erg~cm^{-2}~s^{-1}$ corresponding to a 
$SFR\approx$0.02 $\rm M_{\odot}~yr^{-1}$ (not corrected for extinction).
No continuum is detected above a flux limit $f\approx3\times10^{-19}~\rm erg~cm^{-2}~s^{-1}~A^{-1}$, implying a H$\alpha$ 
equivalent width  $EW(H\alpha)\geq50~\rm\AA$. 
The lower limit on $EW(H\alpha)$ corresponds to an age of the knots $t\leq$5 Myr for an instantaneous burst \citep{starburst}.
This value is significantly shorter than the one obtained from the optical colors (see Fig. \ref{ccdiagram}) implying a significant  
larger amount of dust in the knots ($A_{V}\approx$0.5-1 mag) than the one observed in their parent galaxy ($A_{V}\approx$0.31 mag).
In comparison, for a continuous star formation history the value of $EW(H\alpha)$ corresponds to an age 
$t\leq$1000 Myr, consistent with the estimate obtained from the optical colors, making this scenario much more likely.\\

\subsection{Abell 1689}
\begin{figure*}
\centering
\includegraphics[width=9.5cm]{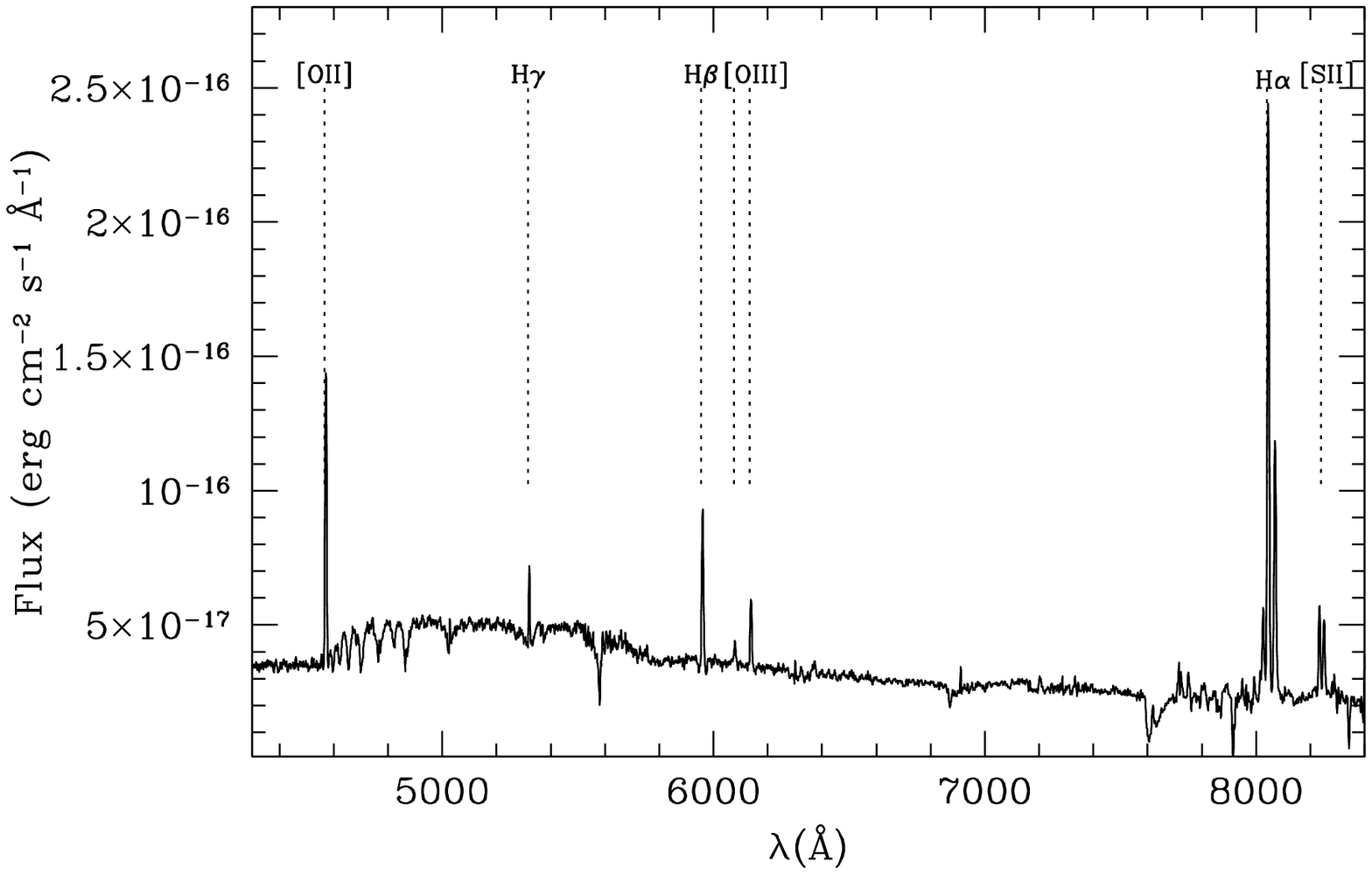}
\includegraphics[width=9.5cm]{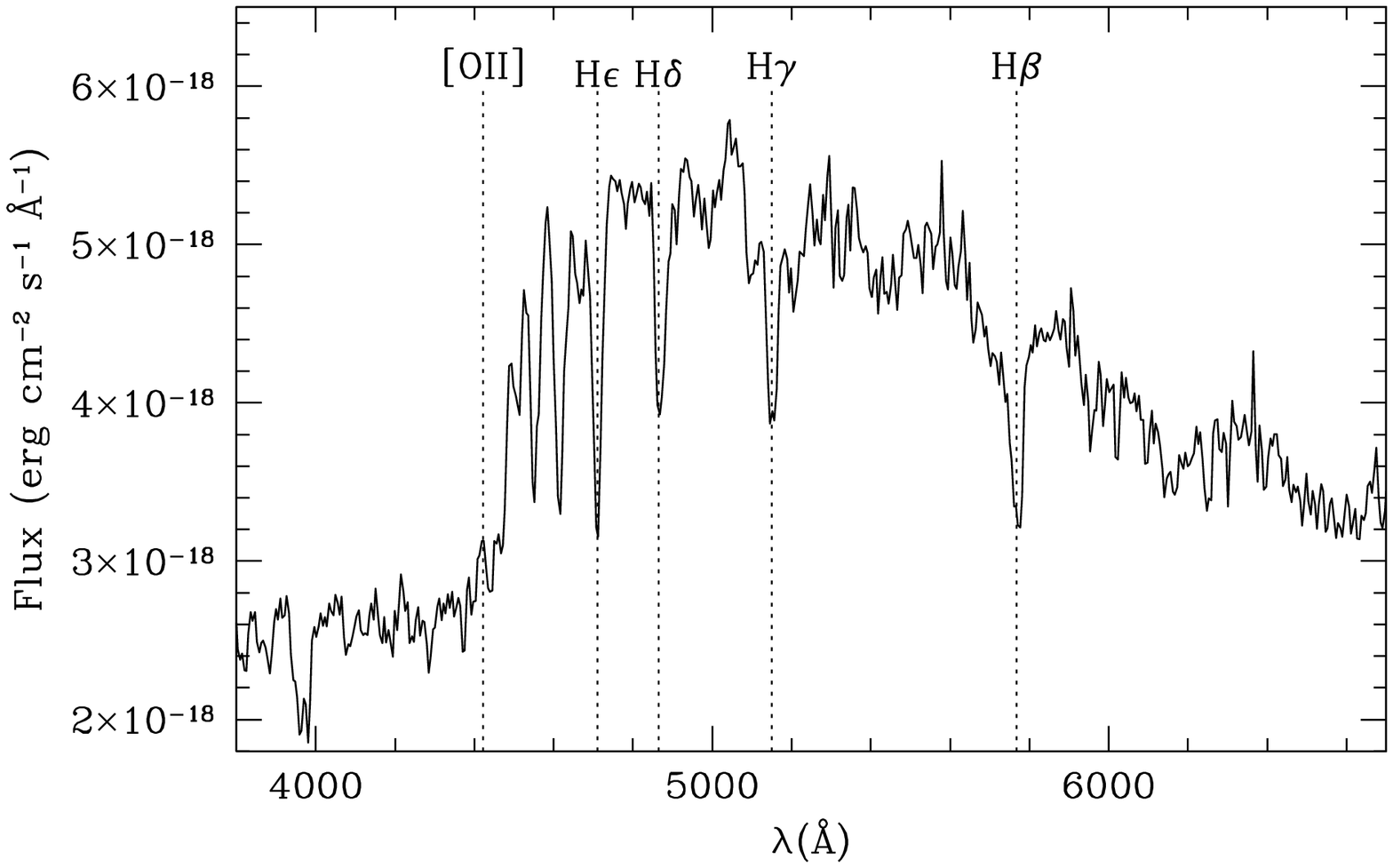}
\caption{The optical spectrum of  235144-260358 in Abell 2667 (upper) and 131124-012040 in Abell 1689 (lower).
The main emission and absorption lines are indicated.}
\label{spectrum}
\end{figure*}

\begin{figure*}
\centering
\includegraphics[width=8.5cm]{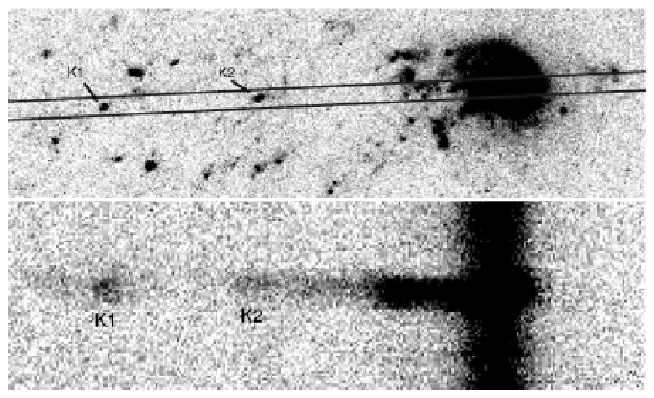}
\includegraphics[width=8.5cm]{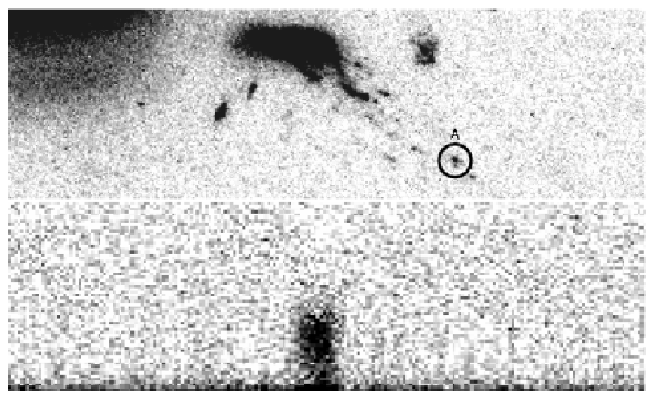}
\caption{Left: The 2D spectrum of 234144-260358 in Abell 2667 and 
of two blue knots near the wavelength of the [OII] emission line (lower panel). 
The slit position is indicated by the two solid lines in the upper panel. [OII] 
diffuse emission appears clearly between the galaxy and the blue knots. Right: 
The 2D spectrum of knot A in Abell 1689 near the wavelength of the [OII] 
emission line (lower panel). The position of the knot is indicated by the black 
circle in the upper panel.}
\label{specknots}
\end{figure*}

\begin{figure*}
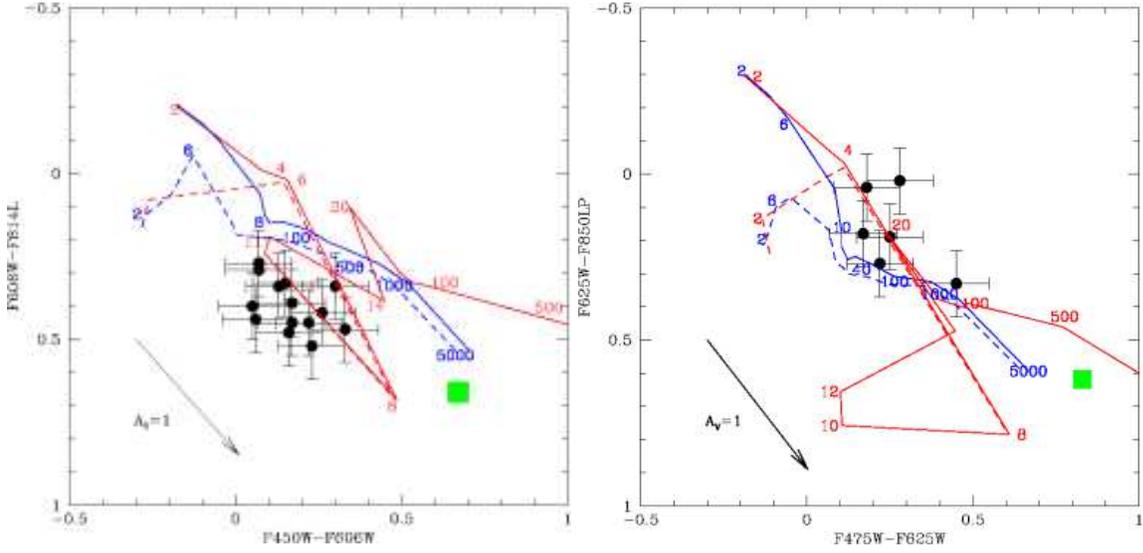

\centering
\includegraphics[width=7.5cm]{A2667color.epsi}
\includegraphics[width=7.5cm]{A1689colors.epsi}
\caption{Colour-Colour diagram for the blue knots in Abell 2667 (left) and Abell 1689 (right).
The black circles with their associated error bars represent the knot colors and their uncertainties.
The green squares indicate the colours of the two disrupted galaxies. The dashed lines are the model predictions for the evolution of 
SSPs (red) and continuous (blue) star formation accounting for the contribution of emission lines. The 
solid lines represent the same models without the contribution of emission lines. The numbers indicate the age of each model in Myr. 
The black arrows show the effect of internal extinction on the observed colours. }
\label{ccdiagram}
\end{figure*}

\begin{figure*}
\centering
\includegraphics[width=7.5cm]{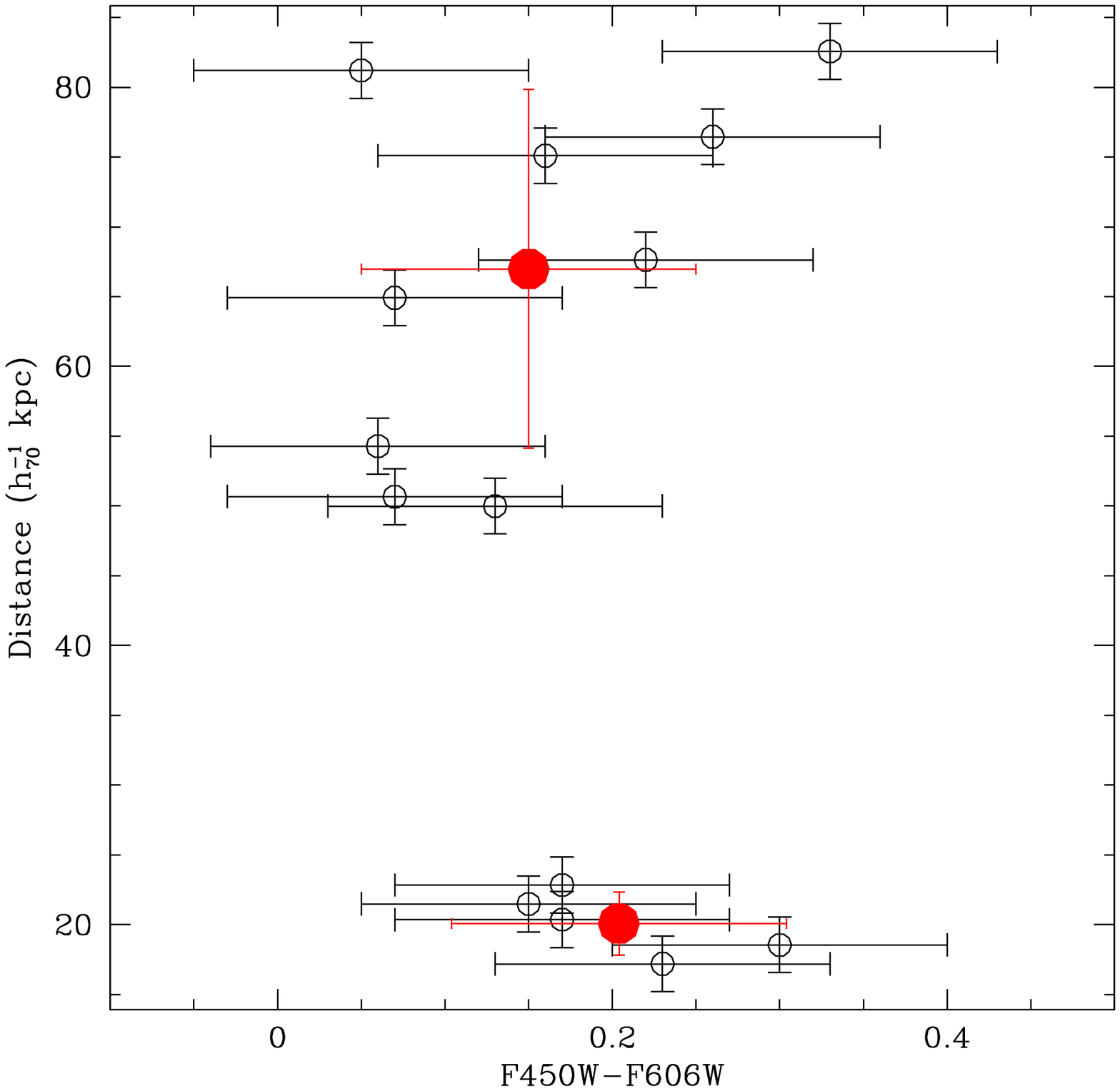}
\includegraphics[width=7.5cm]{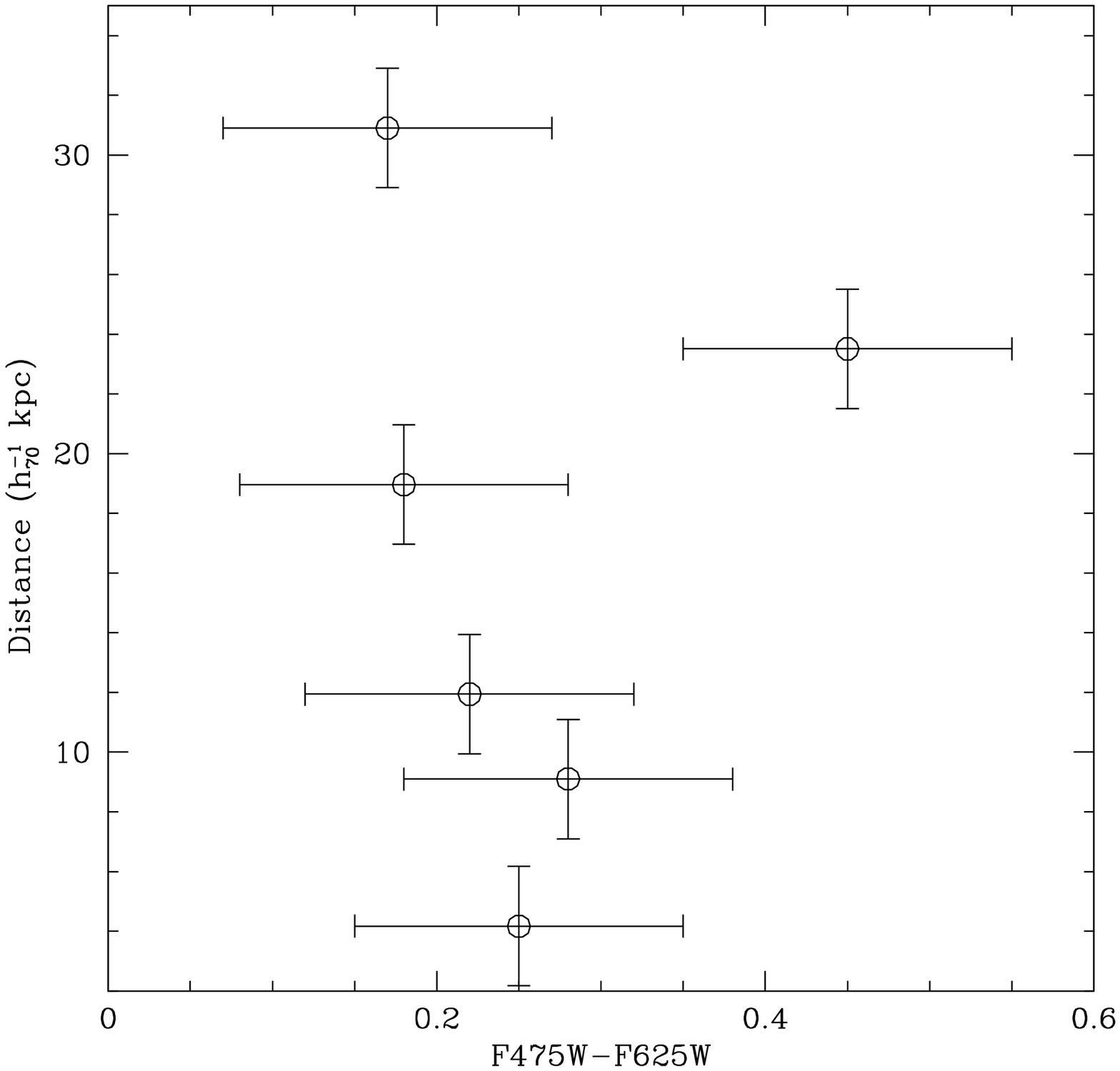}
\caption{Left: Distribution of the F450W-F606W colours for the star forming knots in Abell 2667 
as a function of the distance from 235144-260358. The red points represent the 
average values for the nearest and most distant star forming knots. Right: The same diagram for the F475W-F650W color 
of the star forming knots associated with 131124-012040 in Abell 1689. }
\label{coldistance}
\end{figure*}

The disrupted galaxy in Abell 1689 is illustrated in Fig. \ref{colimage} (lower panel) and 
its main properties are listed in Table 1.
This galaxy lies at a projected distance $\approx$ 0.24 $h_{70}^{-1}$ Mpc from the 
cluster center (see Fig. \ref{whole1689}) and is $\approx$ 2.5 mag fainter than the perturbed galaxy in Abell 2667
(i.e. with a luminosity of $\approx$ 0.1 L$^{*}$, \citealp{wilson97}).
Its redshift is $z\approx$0.1866 confirming that it belongs to A1689.
Contrary to 235144-260358, the surface brightness profile of this galaxy 
follows a typical exponential profile (see Fig.\ref{colprofiles}). However, the slopes of
its color profiles are anomalous: in both $F450W-F625W$ and $F625W-F814W$ there is  
an inversion of the color gradients, with bluer colors toward the center. 
The galaxy outskirts have a $F450W-F814W$ color $\approx1.7$ mag, $\approx$0.6 mag redder than the galaxy center and 
consistent with the typical color of red sequence galaxies in the local Universe \citep{bernar03}.  
Similar features have been observed in spiral galaxies in the Virgo cluster and suggest recent ($t\leq$300 Myr) gas stripping 
by ram pressure \citep{n4569}.
131124-012040 is neither detected at 24 $\mu$m by Spitzer nor in VLA continuum images (see Table \ref{tabgal}).
This is consistent with the optical spectrum of this galaxy (see Fig.\ref{spectrum}), 
which shows strong Balmer lines in absorption 
($EW(H\delta)\approx$6 $\rm \AA$, $D(4000)\approx$1.21) and very little residual 
star formation ($EW([OII])\approx$ 1.8 $\rm \AA$).  
This overall behavior suggests that the galaxy center has 
recently ($t\leq$100 Myr, i.e. \citealp{poggia97,shioya02, kauff03}) stopped forming stars.
These spectral features are consistent with both a simple truncated and a post-starburst 
SFH \citep{shioya02,pracy05,crowl06}, however the inverted color gradients and the absence of a 
central enhancement in the surface brightness profile favor a ram pressure scenario \citep{bekki05ea,n4569}.\\
A $\approx$ 30 $h_{70}^{-1}$ kpc long trail, formed of at least six blue knots and
a number of wisps and filaments, is associated with this 
system. The bright knots have absolute F475W magnitudes in the range -13.5$<M_{F475W}<$-11.5 mag, 
lying between dwarf galaxies and 
stellar clusters ($\approx$ 3 mag fainter than the knots observed in Abell 2667\footnote{ACS observation of A1689 are $\approx$3 mag deeper than WFPC2 imaging of A2667. We cannot therefore exclude that knots as faint as the one in A1689 are present also in A2667.}).
The knots nearest to the galaxy are clearly resolved in the HST images and have 
a typical size $r_{e}(F475W)\approx$0.8-0.9 kpc. In comparison, the most distant knots are not 
resolved implying a physical size $r_{e}(F475W)\leq$0.35 kpc. 
To determine the ages of the blue knots we computed the time evolution of 
the $F475W-F625W$ and $F625W-F850LP$ colors, as described in the previous section.
The results of our analysis are presented in Fig. \ref{ccdiagram}.
Most of the knots lie within the modeled tracks for 
an instantaneous burst with an age in the range 5$<t<$100 Myr, and are slightly above the model for  
continuous star formation with an age in the range 10$<t<$1000 Myr.
As for 235144-260358 no correlation is observed between the optical colors of the knots and their 
distance from the infalling galaxy (see Fig.\ref{coldistance}).
The optical spectrum obtained for the most distant knot (Knot A in Fig.\ref{specknots}), reveals the presence of strong 
[OII] in emission ($f\approx5.4\times10^{-17}~\rm erg~cm^{-2}~s^{-1}~\AA^{-1}$), 
while no continuum is detected at a limit $\approx5\times10^{-19}~\rm erg~cm^{-2}~s^{-1}~\AA^{-1}$, 
implying an $EW[OII]\geq108$ \AA\ and showing that star formation 
is still taking place in this system.
Also H$\alpha$ emission is detected, but it lies on a bright sky line  and is affected by fringing 
making it impossible to use the H$\alpha$ equivalent width to obtain an independent estimate of the age of the burst.
It is interesting to note that the time scale necessary to invert the color gradients ($t\leq$300 Myr) appears to be slightly 
longer than the age of the trails ($t<$100 Myr), suggesting that the two features could be signatures of different physical mechanisms.

\section{Environmental effects on the evolution of the infalling galaxies}
The peculiar properties of the two galaxies falling into A2667 and A1689 suggest 
that both galaxies are undergoing strong transformations due to 
their interaction with the harsh cluster environment. 
However while these objects are at similar distances from the cluster centers and 
show similar extended trails of star-forming 
knots, their recent star formation histories are 
different. 235144-260358 is experiencing a strong burst of star formation, appearing as a rare 
 example of a luminous infrared cluster galaxy.
 In comparison, 131124-012040  has recently ($t\leq$ 100 Myr) ceased 
 its star formation activity.
 To probe this difference, we investigate the 
 effects of different environmental mechanisms on the properties and star formation history of these two galaxies.\\
 The high velocity dispersion of the two clusters ($\sigma_{1D}\geq$1000$\rm km~s^{-1}$; \citealp{covone05,a1689dinam}) makes very unlikely 
 a low velocity interaction or a merger with another cluster galaxy.
This would not be the case if the two galaxies belong to smaller, kinematically distinct, dynamical units (e.g. infalling groups). 
However no observational evidences support this possibility.  
Therefore we will only consider high velocity galaxy-galaxy and galaxy-cluster gravitational interactions  
and ram pressure stripping by the hot intracluster medium (ICM) as possible mechanisms 
to explain the peculiarities of these two galaxies.\\
 
In order to reduce the number of free parameters in our model we assume that 
the two galaxies are falling on linear orbits into the cluster core. 
This very simple scenario, supported by the fact that infalling galaxies 
have usually highly eccentric radial orbits \citep{review}, allow us to express the cluster-centric distance (r) 
as a function of the galaxy infalling velocity:
\begin{equation}
r = \frac{r_{proj}}{sin(arcos \big(\frac{V_{ls}}{V_{infall}}\big))} 
\end{equation}  
where $V_{ls}$ is the (measured) velocity component along the line of sight and 
$r_{proj}$ is the cluster-centric distance projected along the line of sight. 
Similarly, assuming that the trails of blue knots trace the galaxy's trajectory \citep{moore1999}, their physical length is:
\begin{equation}
L_{trail} = \frac{L_{proj}}{sin(arcos \big(\frac{V_{ls}}{V_{infall}}\big))} 
\end{equation}
where $L_{proj}$ is their projected length and their age is: 
\begin{equation}
t_{trails} = \frac{L_{trail}}{V_{infall}} 
\end{equation} 
This value of $t_{trails}$ is based on the assumption 
that the trails are at rest with respect to the cluster, and must be considered as a lower limit for the real 
age of these features.\\
Both clusters have a 1D velocity dispersion 
$\sigma_{1D}\geq1000\rm km~{s}^{-1}$ \citep{covone05,a1689dinam}, implying a 3D 
infalling velocity $V_{infall}\approx\sqrt{3}\sigma_{1D}$.
In the following we therefore assume a 3D 
infalling velocity in the range between 1000$<V_{infall}<$1730 $\rm km~{s}^{-1}$ (i.e. $\sqrt{3}\sigma_{1D}$ and  $\sigma_{1D}$) that can be considered as upper and lower limit of the real value.
The values so derived for the cluster-centric distance, the length and the age of the trails for the upper and lower 
limit of $V_{infall}$ are summarized in Table \ref{model_assumption}.

\subsection{Gravitational interactions}
\begin{figure*}
\centering
\includegraphics[width=9cm]{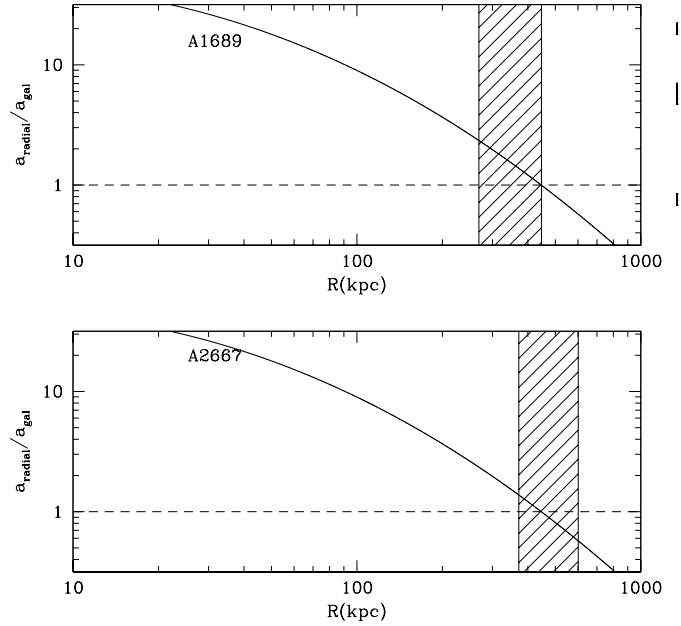}
\caption{The ratio of the radial acceleration due to the cluster potential to the internal galaxy acceleration as a function 
of the cluster-centric distance for A1689 (upper panel) and A2667 (lower panel). The shaded regions indicate the range 
of cluster-centric distances assumed in the model (see Section 4). }
\label{accrad}
\end{figure*}

\begin{figure*}
\centering
\includegraphics[width=7.5cm]{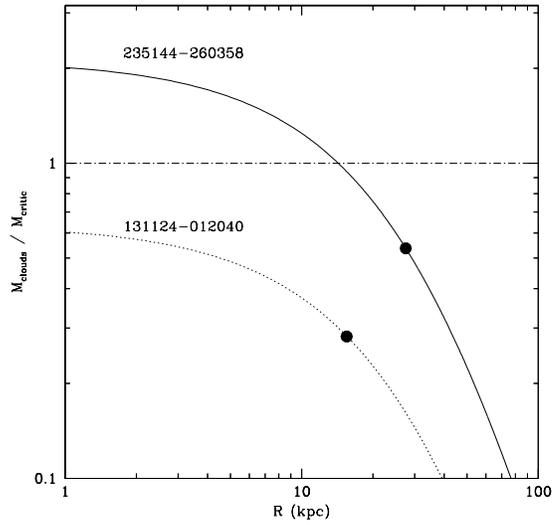}
\caption{The ratio of the typical HI cloud mass in galactic disks to the critical mass for gas collapse in the case of galaxy galaxy interactions as a function of the distance from the perturber for  235144-260358 in Abell 2667 (solid line) and of 131124-012040 in Abell 1689 (dotted line). The black circles indicate the values for the closest interaction possible (r$\approx R_{gal}$).
The horizontal dotted-dashed line shows the transition between the stable ($M_{clouds}/M_{critic}<$1) and unstable ($M_{clouds}/M_{critic}>$1) regime. }
\label{highvel_mcrit}
\end{figure*}

\begin{figure*}
\centering
\includegraphics[width=9cm]{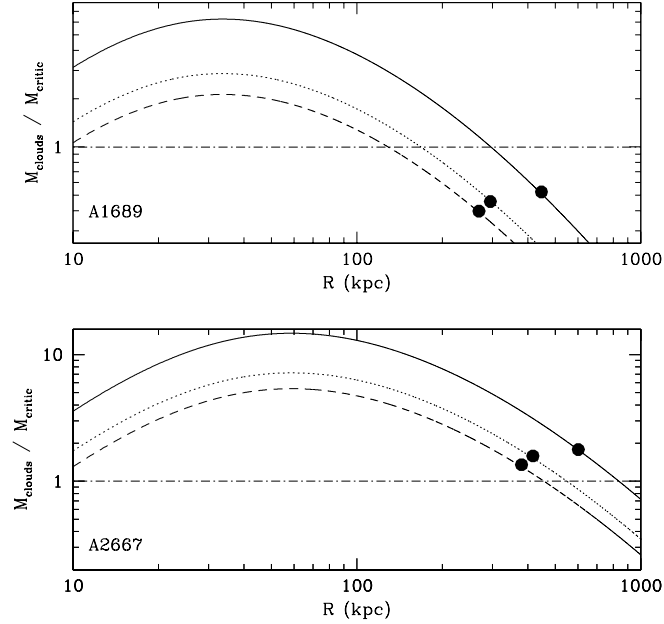}
\caption{Same as Fig.\ref{highvel_mcrit} in case of a cluster-galaxy interaction in A1689 (upper panel) and A2667 (lower panel), for three different values of the 3D infalling velocity: 
1000$\rm ~km~s^{-1}$ (solid line), 1410$\rm~ km~s^{-1}$ (dotted line) and 1730$\rm ~km~s^{-1}$ (dashed line). The black circles indicate the current galaxies positions assumed in each model.}
\label{mcrit}
\end{figure*}

We can approximate the strengths of high velocity galaxy-galaxy and galaxy-cluster interactions 
by using the impulse approximation  \citep{byrd96}.
The transverse and radial tidal accelerations 
experienced by the infalling galaxy are:
\begin{equation}
a_{tr} = GM_{pert}\frac{R}{[R^{2}+(r+R)^{2}]^{1.5}}
\end{equation}  
\begin{equation}
a_{rad} = GM_{pert}\big[\frac{1}{r^{2}} - \frac{1}{(r+R)^{2}}\big]
\end{equation}
where $M_{pert}$ is the mass of the perturber within $r$, 
$R$ is the radius of the perturbed galaxy (assumed to be $\approx$5 effective radii \citep{gav00}) and $r$ is 
its distance from the perturber.
The radial tidal field tends to accelerate the edge of a galaxy. If it is 
more intense than the internal galaxy acceleration, given by 
\begin{equation}
a_{gal} = \frac{GM_{dyn}}{R^{2}}
\end{equation}  
where $M_{dyn}$ is the dynamical galaxy mass, it is able to strip material from the infalling galaxy.
Following \cite{phenomen}, we use the H-band rest frame luminosity of the two 
galaxies to estimate their dynamical masses within the optical radius  and derive their disk rotational velocities obtaining $M_{dyn}\approx10^{11.6}~\rm M_{\odot}$ and $M_{dyn}\approx10^{10.6}~\rm M_{\odot}$ for 235144-260358 and 131124-012040 respectively.
In the case of non-interpenetrating galaxy-galaxy interactions, the impact parameter is at least equal or greater than 
the galactic radius (i.e. $r\geq R$) implying that material is stripped from an infalling galaxy only if 
\begin{equation}
\label{highvel}
M_{pert}\geq1.33\times M_{dyn}
\end{equation} 
(i.e. $\approx10^{11.7} \rm M_{\odot}$ and $\approx10^{10.7} \rm M_{\odot}$ for 235144-260358 and 131124-012040 respectively). 
The possible perturber should not lie at a larger distance than the typical size of trails.
In Abell 2667, the brightest objects within a projected distance of 100 kpc from 235144-260358 have an H band magnitude 
$\approx$ -24.5 mag (i.e. $M_{dyn}\approx10^{11.6}~\rm M_{\odot}$), fairly consistent with the lower limit required 
for effective stripping. Unfortunately their recessional velocities are unknown, making impossible a more detailed analysis of their 
possible interaction with 235144-260358.   
In Abell1689, the giant face-on barred spiral projected at $\approx$20 kpc NE from 131124-012040  (see Fig. \ref{whole1689}) is the only object (within 100 kpc) 
satisfying Equation \ref{highvel}. 
However the trail of blue knots is pointing in the opposite direction than the one expected in the case of interaction between the two objects (i.e. towards the perturber) and the galaxy has a redshift of $z\approx$0.1924 i.e. 1680 km/s higher than the recessional velocity of 131124-012040, making unlikely an interaction between the two objects.\\
To quantify the effect of tidal forces from the cluster potential well on an infalling galaxy, 
we assume a NFW profile \citep{NFW} for the cluster mass distribution:
\begin{equation}
M (<r) = M_{0}  [ln(1+\frac{r}{r_{s}}) - \frac{r/r_{s}}{1+r/r_{s}}] ~~~{\rm for}~r\leq r_{s}c
\end{equation}  
where
\begin{equation}
M_{0} = 4\pi  \frac{3H_{0}^{2}}{8\pi G}\big[\Omega_{M}(1+z)^{3}+\Omega_{\Lambda}\big] \frac{200 c^{3} r_{s}^{3}}{3 [ln(1+c)-c/(1+c)]}
\end{equation}  
and $r_{s}$ and $c$ are the scale radius and concentration parameter of the mass distribution. The values 
adopted for the two clusters are summarized in Table \ref{cluster}.
As shown in Fig.\ref{accrad}, for both our galaxies the radial acceleration from the cluster potential is higher than the 
internal acceleration for a cluster-centric distance smaller than $\approx$ 0.45 $h_{70}^{-1}$ Mpc. 
Therefore, depending on their real infalling velocity, the two objects are at the edge or are just entered the region 
where material can be efficiently stripped by gravitational interactions.
This simple calculation gives a lower limit for the real efficiency of mass loss,  
since higher rates will occur in the presence of substructures and infalling groups, as is likely in these 
two clusters \citep{covone06b,a1689dinam}. Moreover, tidal heating \citep{taylor01} produced by the varying 
cluster gravitational field will significantly accelerate mass loss \citep{gnedin03b,gnedin03}, 
although it is not considered in our model.\\ 
In contrast to the radial acceleration, which tends to strip material from an infalling galaxy, 
the transverse field compresses the 
interstellar medium to produce gas infall toward the center and may trigger a burst of star formation. 
Gas clouds experience a velocity perturbation due to the transversal tidal acceleration 
and collide with other gas clouds.
The increase in the cloud velocity can be estimated as:
\begin{equation}
V = \int a_{tr} dt \approx a_{tr} \Delta(t)
\end{equation}
where $\Delta(t)$ is the age of the interaction.
The velocity increase and cloud collision produce a density enhancement in the centre of the galaxy, which is 
proportional to the Mach number ($MN$) squared. 
Consequently the critical mass for the cloud collapse 
(which is proportional to $\rho_{gas}^{-0.5}$) decreases by a factor $MN^{-1}$ and, in the case of a strong perturbation, could become 
smaller than the typical mass of a galactic disk HI cloud ($\approx$300 $\rm M_{\odot}$, \citealp{spitzer78,jog92}), 
favoring new episodes of star formation  \citep{byrd96}.
Fig. \ref{highvel_mcrit} shows the ratio between the typical mass of HI clouds and the critical mass for cloud 
collapse in the case of high velocity interactions, assuming $M_{pert}\approx10^{11.6} \rm M_{\odot}$ and  $M_{pert}\approx10^{11.3} \rm M_{\odot}$ 
for 235144-260358 and 131124-012040 respectively as discussed above.
It appears that in both cases high velocity galaxy-galaxy interactions 
are not enough strong to trigger a burst of star formation as observed in 235144-260358.\\
This is not the case of galaxy-cluster interactions.
Fig. \ref{mcrit} shows again the ratio between the typical mass of HI clouds and the critical mass for cloud 
collapse as a function of the cluster-centric distance in case of interaction with the cluster potential. 
In this case two galaxies are in two different regimes, whatever are the initial conditions in our model.
While in 235144-260358 the critical mass is already below  $\approx$300 $\rm M_{\odot}$ and the 
compressed gas is able to collapse and produce new stars, in 131124-012040 this is still not the case.
This result is consistent with our observations and 
indicates that tidal forces from A2667 
may have triggered the strong starburst in 235144-260358.\\
In summary our model suggests that gravitational interactions with the cluster potential alone are able 
to strip material from the two infalling galaxies and to trigger a burst of star formation in 235144-260358.
Even if we cannot completely exclude a role of high velocity galaxy-galaxy interactions on the evolution 
of these systems, it appears clear that they cannot account for all the properties of the two infalling galaxies.

\subsection{Ram pressure stripping}
\begin{figure*}
\centering
\includegraphics[width=8cm]{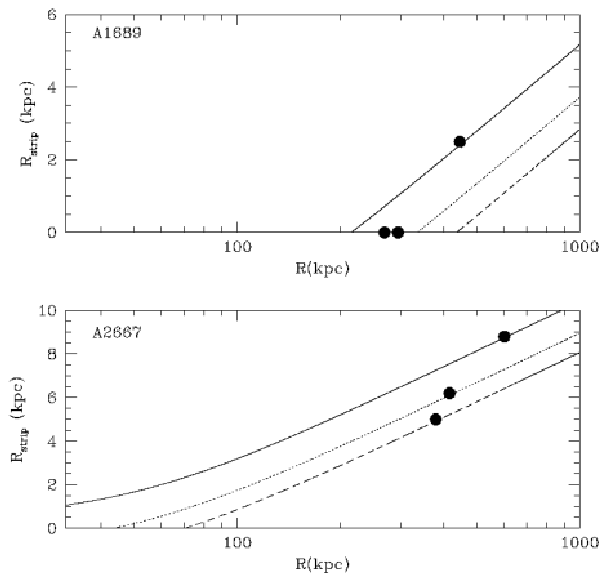}
\includegraphics[width=8cm]{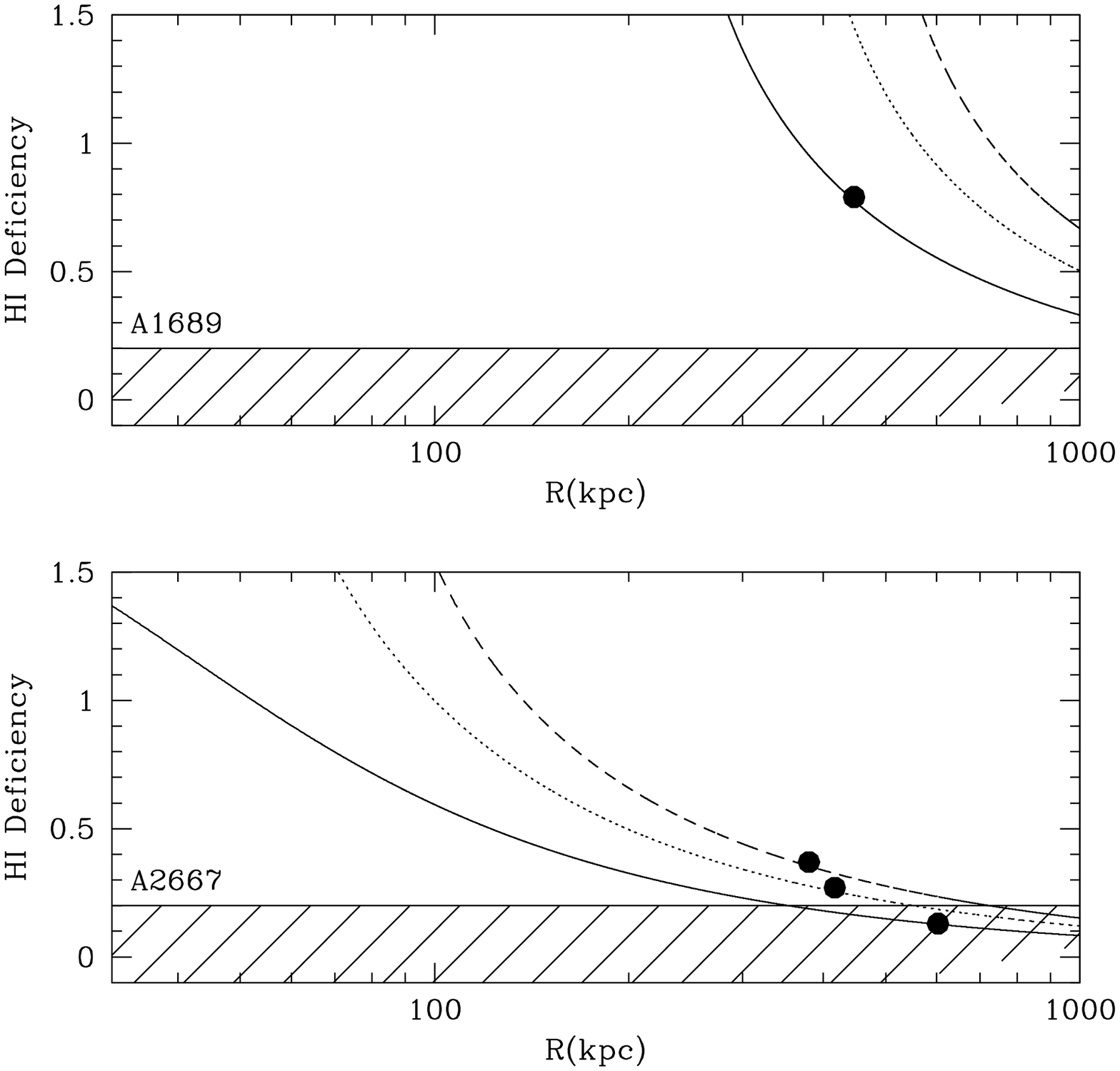}
\caption{Left: The variation of the stripped radius as a function of the cluster-centric distance for 235144-260358 in Abell 2667 (lower panel) and of 131124-012040 in Abell 1689 (upper panel) for three different values of the 3D infalling velocity. Symbols as in Fig.\ref{mcrit}. Right: The variation of the HI deficiency as a function of the cluster-centric distance for 235144-260358 in Abell 2667 (lower panel) and of 131124-012040 in Abell 1689 (upper panel) for three different values of the 3D infalling velocity. The shaded area indicates the typical range of HI-deficiency observed in unperturbed galaxies. Symbols as in Fig.\ref{mcrit}. For $V_{infall}\geq1200~\rm km~s^{-1}$ the galaxy has lost all its original HI content.}
\label{rstrip}
\end{figure*}

Although the tidal interaction hypothesis is consistent with the presence of a strong starburst 
only in 235144-260358, it is not able to explain why 131124-012040 shows clear signs of a 
recent truncation of its star formation. 
Both A1689 and A2667 are X-ray bright clusters suggesting that the effects of the hot intracluster medium 
could be significant.
Therefore, to estimate the effects of ram pressure stripping on the infalling galaxies, we 
adopt the classical \cite{GUNG72} criterion:
\begin{equation}
P_{ram}= \rho _{ICM} v^{2} \geq 2 \pi G \Sigma_{star} \Sigma_{gas}
\end{equation}
where $\rho_{ICM}$ is the density of the cluster medium, $\Sigma_{star}$ and $\Sigma_{gas}$ 
are the galaxy stellar and gas density, and
$v$ is the 3D infalling velocity of the galaxy (here assumed to be in the range 1000$<v<$ 1730 $\rm km~s^{-1}$ 
as discussed in the previous sections).
We use a $\beta$ model density profile for the ICM: 
\begin{equation}
\rho (r) = \rho _{0} \frac{1}{[1+(r/r_{c}^2)]^{3\beta/2}}
\end{equation}
(the values adopted for the different clusters are listed in Table \ref{cluster}).
We assume that the stellar and gas distributions of our galaxies are exponential, 
as confirmed by their structural parameters.
The gas and stellar density profiles are \citep{domainko06}
\begin{equation}
\Sigma_{star,gas} (r) = \frac{M_{star,gas}}{2 \pi R_{0star,gas}^{2}} \exp(-r/R_{0star,gas})
\end{equation}
where $R_{0}$ is the scale length of the exponential profile (i.e. 0.59 $r_{e}$).
Assuming a gas scale length $R_{0gas}\approx1.8~R_{0star}$ \citep{cayatte94}, a 
$M_{gas}/M_{star}$ ratio $\approx$1, typically observed in late type galaxies \citep{boselli}, and a 
$M_{star}/L_{H}$ ratio $\approx$ 1 \citep{phenomen,mcgau00} the 
typical stripping radius is given by the following relation:
\begin{equation}
R_{strip} \approx 0.64 R_{0} \times \ln \big( \frac{G (L_{H}/L_{\odot})^{2}}{1.8^{2}\rho _{ICM} v^{2} 2 \pi R_{0star}^{4}} \big)
\end{equation}
and the mass of gas stripped by ram pressure is:
\begin{equation}
M_{strip} = \frac{L_{H}}{L_{\odot}} (\frac{R_{strip}}{R_{0}} + 1) exp(-R_{strip}/R_{0})
\end{equation}
In Fig. \ref{rstrip} (left) we show the variation of the stripping radius as a function 
of the distance from the cluster center, for three different values of the 3D infalling velocities 
assumed in our model ($\approx$1000$km~s^{-1}$, $\approx$1410$km~s^{-1}$ and $\approx$1730$km~s^{-1}$).
While at its current location 131124-012040 has almost been totally stripped by ram 
pressure ($R_{strip}\leq 0.9 \rm r_e$), in 235144-260358 ram pressure 
has only affected the outer galaxy regions ($R_{strip}\geq 1.2 \rm r_e$). 
The same result can be analyzed in terms of HI deficiency\footnote{The 
HI deficiency is defined as the difference, in logarithmic units, between the 
observed HI mass and the value expected from an isolated galaxy 
with the same morphological type T and optical linear diameter D:
HI DEF = $<\log M_{HI}(T^{obs},D^{obs}_{opt})> - log M^{obs}_{HI}$ \citep{haynes}} (see Fig. \ref{rstrip} right).
131124-012040 has already lost $\geq$80 \% of its original gas content and if observed in 
a local cluster it would be classified 
as a highly HI-deficient object.
Conversely, ram pressure has only stripped a tiny fraction of the gas from 235144-260358, whose HI 
deficiency ($\approx$0.25) would approximately 
lie at the edge between normal and deficient galaxies (HI-deficiency$\approx$ 0.2).
We remark that the HI-deficiency shown in Fig.\ref{rstrip} is not determined from observations, but is 
obtained from our analytical model.\\
Comparing Fig.\ref{rstrip} to Fig.\ref{accrad} it appears clear that ram pressure in 131124-012040 has become efficient 
before gravitational interactions were able to strip material from the galaxy. This is qualitatively consistent 
with the different time-scales of the interaction determined from the inversion of the color gradients ($t\leq$ 300 Myr, likely produced by ram pressure 
stripping) and from the blue star-forming trails ($t<$ 100 Myr, clear signature of gravitational interactions).

\subsection{The origin of the blue star forming knots}
The mutual effects of gravitational interactions and ram pressure have already been 
observed in several cluster galaxies, however the tails of blue star forming knots here discovered represent an 
extremely rare feature, to our knowledge previously observed only one other 
starburst galaxy in the Abell cluster 2125 ($z\approx$0.247, \citealp{owen06}).
The morphology and luminosity of the knots suggest that we are dealing with dwarf galaxies 
and/or stellar super-clusters \citep{felhauer02}. 
This could explain the observed difference between luminosity of the knots ($\approx$2.5 mag) in 
the two galaxies since the luminosity of the brightest star clusters is usually correlated with the SFR of the parent galaxy \citep{weidner04}.\\
The properties of the knots (i.e. colors and emission lines) suggest that they 
are undergoing an extended period of star formation as discussed in Section 3.
From the model described in the previous sections the dynamical age of the trails results 
50$<t_{trail}<150$ Myr and 20$<t_{trail}<60$ Myr for 235144-260358 and 
131124-012040 respectively, fairly consistent\footnote{However the value of $t_{trail}$ obtained in the previous section must be considered as 
a lower limit since it assumes that the stripped material is at rest with regard to the cluster.} 
with the age inferred from their optical colors 
(see Fig. \ref{ccdiagram}).  
It is impossible to determine whether the knots 
where already forming stars when they have been stripped or 
if their activity was triggered by an external mechanism once in the ICM. 
We can exclude that these systems have been formed in the ICM by the accretion of 
unbounded material stripped by the parent galaxies: the combined effects of the cluster tidal field and 
ram pressure tend to inhibit the formation of bound systems from the collapse of stripped material \citep{mihos04}.
This scenario is consistent with the  numerical simulations.
\cite{elmegreen} showed that gravitational interactions can lead to the formation and ejection of peripheral self 
gravitationally bound clouds with masses $\leq10^{8}~\rm M_{\odot}$, which begin their 
life in a major burst of star formation. 
Moreover \cite{bekki03} have recently demonstrated that ram pressure can trigger 
the collapse of stripped clouds leading to a burst of star formation, suggesting that even the formation and evolution 
of the blue star forming knots is probably driven by the mutual effects of gravitational interactions and ram pressure.
Only deeper spectroscopic observations will shed light on the star formation history of these rare objects.

\section{Discussion \& Conclusion}
The analysis in this paper allows us to propose a scenario 
for the evolution of the two disturbed galaxies in Abell 2667 and Abell 1689.
These objects are currently falling into massive, gas-rich galaxy clusters with similar mass 
and gas density profiles (see Table \ref{cluster}). 
Under the combined action of tidal forces (more likely from the cluster potential) and of ram pressure 
by the ICM, their morphologies and star formation are strongly perturbed. 
Self gravitational bound systems are ejected 
from the main galaxies and stars and ionized gas are stripped from the stellar disks 
producing the observed tails of blue knots and stellar wisps 
tracing the infalling trajectory of these systems into the cluster core. 
Only the  tidal field of Abell 2667 is able to drive a gas infall into the center of 235144-260358 triggering a nuclear burst 
of star formation and making this galaxy a rare example of luminous infrared cluster galaxy. 
Simultaneously ram pressure stripping by the hot intracluster medium strips the neutral hydrogen from 
the galaxy outskirts but it is not able to affect the central regions where the starburst is taking place. 
Conversely in 131124-012040 gravitational forces are not strong enough to trigger 
the collapse of gas clouds while ram pressure is already extremely efficient.
At the present galaxy location ram pressure has stripped at least the $\approx$80\% of the original neutral hydrogen content, 
quenching the star formation activity in this object, as confirmed by the strong Balmer line 
in absorption observed in the optical spectrum \citep{shioya02} and by the inversion of the optical color gradients 
along all the galaxy extent \citep{review,n4569}.\\
A larger statistical sample is necessary to determine whether we are witnessing 
a common snapshot in the evolution of cluster galaxies or an extremely rare phenomenon. 
In fact, as discussed in the Introduction, only these 2 galaxies out of 13 different clusters imaged 
at 0.175$<z<$0.25 show extended trails of blue knots.
Within the WFPC2 field of view ($\approx$0.25 $\rm Mpc^{2}$ at $z\approx0.2$) there are typically $\approx$50 
cluster members but only the $\approx$20\% of them are spiral galaxies \citep{balogh02}, 
implying a frequency of $\approx$1.5\% (2 over 130) among cluster spirals at $z\approx$0.2.
This value is fairly consistent with the expected frequency roughly obtained dividing the typical 
time scale of the interaction ($\leq$200 Myr) to the age of the cluster ($\approx$11 Gyr).\\
If we are witnessing a common step of cluster galaxy evolution what can we learn by studying these two rare objects?
Abell 2667 and Abell 1689 have comparable mass and gas profiles and the two galaxies are approximately 
at the same projected cluster-centric distance, suggesting that the absolute intensity of the cluster tidal field and of ram pressure 
by the cluster ICM is approximately the same in the two environments.
Therefore we can speculate that the different recent evolutionary history of the infalling systems could be in part 
related to their different properties (i.e. their different luminosities: $\approx\rm  L^{*}$ and $\approx$0.1L$^{*}$).
In this case, our result suggests that giant spiral galaxies infalling into the core of massive clusters are 
mainly perturbed by the gravitational interaction with the cluster. 
Stars respond by forming arms and bars, while the gas flows directly toward the central region within $t\approx$100 Myr.
The sinking of the gas towards the center triggers a burst of star formation and is able to alter the galaxy 
morphology (\citealp{iono}). Ram pressure stripping produces a truncation of the disk but only in the outskirts of the 
galaxy being not efficient within the optical effective radius. 
When all the remaining fuel has been consumed by the star formation this galaxy will not longer
appear as a disky gas rich systems but more likely as a bulge dominated quiescent spiral.
This is not the case for less massive galaxies. Ram pressure is much more efficient on low mass systems and it is able to 
strip a considerable fraction of the neutral hydrogen from the inner part of these galaxies, preventing the gas from 
sinking toward the center driven by tidal interaction and quenching the star formation history. 
Within $\approx$1 Gyr this object will not appear as a blue spiral any more but 
will probably look like an early type (e.g. red) disky spiral \citep{shioya02}.
The different evolutionary scenarios for the evolution of low and high mass infalling galaxies emerging from our analysis 
apparently fit with recent observations and models, suggesting that the bulk of the cluster population of giant bulge 
dominated early type spiral galaxies can only be formed during some kind of gravitational 
interaction \citep{dress04car,mihos04}, while lower mass systems can be transformed by simple gas 
removal from healthy spirals \citep{poggianticoma}.\\
The properties of the blue knots stripped from the infalling galaxies deserve particular attention. 
These systems have a  luminosity (-16.5$\leq M \leq$-11.5) and 
a physical size ($r_{e}\leq0.45$ kpc) typical of dwarf galaxies and consistent with the ultra compact dwarf 
galaxies (UCD, \citealp{hinker99,philips01}), recently discovered in Abell 1689 \citep{mieske04,mieske05}.
There are two competing formation scenarios to explain the origin of UCDs.
\cite{bekki03b} propose that they are the remnants of stripped dwarf galaxies.     
In this scenario a nucleated dwarf looses its envelope and a great part of its dark matter content due to tidal interaction 
with another object. On the contrary \cite{felhauer02} propose that UCDa could origin from the \emph{amalgamation} 
of rich aggregates of young massive star clusters that can form during gravitational interactions between gas-rich galaxies.
It appears clear that the knots discovered here strongly support the second scenario, suggesting that at least part 
of the population of ultra-compact dwarfs originate from young massive star clusters: we are probably 
for the first time witnessing the dawn of the UCDs.
This scenario is also consistent with the recent discovery of a massive extra galactic 
star cluster ($M\geq10^{6}~ \rm M_{\odot}$, $t\approx700$ Myr) lying at a projected distance of 17 kpc from 
the merger remnant NGC3310 and likely formed during the merging event \citep{knapp06}.\\
Finally the diffuse stellar streams and ionized gas observed along the trails 
suggest that the mechanisms acting here
will significantly influence the properties of the intracluster light and contribute 
to the enrichment of the ICM.
The results here obtained might be representative only of the clusters at $z\geq$0.2 where the infalling rate 
is higher and galaxies have an higher gas content than the one observed in local clusters of galaxies.

\section*{Acknowledgments}
We thank the referee, D. Christlein, for his useful comments which helped us to improve and strengthen the paper. 
LC is supported by the U.K. Particle Physics and Astronomy Research Council.
Part of the data presented herein were obtained at the W.M. Keck Observatory, which is operated as a scientific partnership among the
California Institute of Technology, the University of California and
the National Aeronautics and Space Administration. The Observatory was
made possible by the generous financial support of the W.M. Keck
Foundation.This work was partially supported by NASA 
contract 1255094, administered by JPL/Caltech.
JR acknowledges support from Caltech

\newpage

\begin{table*}
\caption{Properties of the two disrupted galaxies}             
\label{tabgal}      
\centering                          
\begin{tabular}{lcclc}        
\hline\hline                 
                          	    	  &  A1689	     &  \vline  &                                 	     & A2667	         \\
                          	    	  &  $131124-012040$ &  \vline 	&                                 	     & $235144-260358$    \\
\hline                        
${\alpha}$ (J.2000)               	  &  13:11:24.86     &	\vline & ${\alpha}$ (J.2000)	          	     &   23:51:44.03      \\
${\delta}$ (J.2000)               	  &  -01:20:39.9     &	\vline & ${\delta}$ (J.2000	          	     &  -26:03:59.6        \\
$z$					  & 0.1870             &  \vline & $z$     &0.2265  \\
$F475W$	  (mag)		    	  	  &    -19.09	     &	\vline & $F450W$  (mag) 		  	     &  -21.54              \\
$F625W$	  (mag)		    	  	  &    -19.95	     &	\vline &  $F606W$  (mag)		  	     &   -22.20             \\
$F850LP$	  (mag)	    	  	  &    -20.49	     &	\vline &  $F814W$  (mag)		  	     &   -22.87            \\
$H$	(mag)	            	  	  &    -21.82	 &	\vline &  $H$  	(mag)			  	     &   -24.47            \\
$F(3.6\mu m)$  (mJy)                 &  0.026    &  \vline & $F(3.6\mu m)$  (mJy)               &  0.41  \\
$F(4.5\mu m)$  (mJy)                 &  0.019    &  \vline & $F(4.5\mu m)$  (mJy)               &  0.43  \\
$F(5.8\mu m)$  (mJy)                 &  0.005    &  \vline & $F(5.8\mu m)$  (mJy)               &  0.29  \\
$F(8\mu m)$  (mJy)                     &  0.008    &  \vline & $F(8\mu m)$  (mJy)               &  1.81  \\
$F(24\mu m)$  (mJy)               & $<$0.03    & \vline &   $F(24\mu m)$  (mJy)               &  4.20  \\
$F(20 cm)$  (mJy)               & $<$0.90    & \vline &   $F(20 cm)$  (mJy)               &  1.40  \\
$r_{e}$	(F850LP) ($h_{70}^{-1}$ kpc)      &   3.1              &  \vline &  $r_{e}$     (F814W) ($h_{70}^{-1}$ kpc)    &     5.5             \\
$\mu_{e}$ (F850LP) (mag $\rm arcsec^{-2}$)&   20.07          &  \vline &  $\mu_{e}$   (F814W) (mag $\rm arcsec^{-2}$)& 19.99               \\
$C_{31}$ (F850LP)                         &   3.13           &  \vline &  $C_{31}$    (F814W)	                     &	   3.02           \\
\hline                                   
\end{tabular}
\end{table*}

\begin{table*}
\caption{Cluster centric distances and properties of the trails as a function of the 3D infalling velocity assumed in the model described in Section 4.}             
\label{model_assumption}      
\centering                          
\begin{tabular}{lccccccc}        
\hline\hline                 
   Cluster    ~~~~~&  \multicolumn{3}{c}{$V_{infall}\approx$1730 $\rm km~s^{-1}$}	 &   ~~~~~ &   \multicolumn{3}{c}{$V_{infall}\approx$1000 $\rm km~s^{-1}$}	         \\
                          	    	  & D  & $L_{trails}$  	& $ t_{trails}$                        &    	     & D  & $L_{trails}$  	&  $t_{trails}$                         \\
		                   & kpc  & kpc               &   Myr                                       &           &  kpc      & kpc     &  Myr    \\
\hline                        
A1689                        &     280      & 35     &       20               &       &     450    &   60        &  60       \\ 
A2667                        &     390      &  90    &       54               &       &     610   &    140     &   150        \\ 
\hline                                   
\end{tabular} 
\end{table*}

\begin{table*}
\caption{Best fitting parameters for the total mass and gas density profiles of A1689 and A2667}             
\label{cluster}      
\centering                          
\begin{tabular}{lcccccccc}        
\hline\hline                 
   Cluster             ~~~~~ 	  &  \multicolumn{3}{c}{Mass Profile}	 &   ~~~~~ &   \multicolumn{4}{c}{Gas Profile}	         \\
                          	    	  & $r_{s}$ & $c$ 	&  Ref.                            &    	     &    $\beta$ & $R_{c}$ & $\rho_{0}$ & Ref.  \\
		                   & Mpc      &      &                                         &         &                  &  Mpc        & g $\rm cm^{-3}$ &   \\
\hline                        
A1689                        &      0.31      & 8.2     &             1               &       &     0.55    &   0.07        &     1.77$\times10^{-24}$        & 3     \\ 
A2667                        &      0.49      &  6    &              2               &       &     0.52   &      0.049   &   2.17$\times10^{-24}$           & 3    \\ 
\hline                                   
\end{tabular}
\newline
References: 1)\cite{broad05} 2) \cite{covone05}; 3) \cite{ota02}. 
\end{table*}

\end{document}